\documentclass[a4paper]{article}
\usepackage[utf8]{inputenc}

\pdfoutput=1
\usepackage{titling}
\usepackage[backend=biber,sortcites=false, style=apa]{biblatex}
\usepackage{booktabs}
\usepackage[table,xcdraw]{xcolor}
\usepackage{setspace}
\usepackage[margin=2.5cm]{geometry}
\usepackage{graphicx}
\usepackage{xcolor}
\usepackage{graphicx}
\usepackage{float}
\usepackage{amssymb,amsmath,amsthm}
\numberwithin{equation}{section}
\usepackage{bbm}
\usepackage{subcaption}
\usepackage{bm}
\usepackage{authblk}

\title{Designing Randomized Experiments to Predict Unit-Specific Treatment Effects}
\author[1]{Elizabeth Tipton}
\author[2]{Michalis Mamakos}
\affil[1]{Department of Statistics and Data Science, Northwestern University}
\affil[2]{Department of Psychology, Northwestern University}
\date{October 17, 2023}
\begin{document}
\maketitle

\begin{abstract}
Typically, a randomized experiment is designed to test a hypothesis about the average treatment effect and sometimes hypotheses about treatment effect variation. The results of such a study may then be used to inform policy and practice for units not in the study. In this paper, we argue that given this use, randomized experiments should instead be designed to predict unit-specific treatment effects in a well-defined population. We then consider how different sampling processes and models affect the bias, variance, and mean squared prediction error of these predictions. The results indicate, for example, that problems of generalizability --- differences between samples and populations --- can greatly affect bias both in predictive models and in measures of error in these models. We also examine when the average treatment effect estimate outperforms unit-specific treatment effect predictive models and implications of this for planning studies.
\end{abstract}

\let\thefootnote\relax\footnotetext{Correspondence concerning this article should be addressed to Michalis Mamakos, Dept. of Psychology, Northwestern University, 2029 Sheridan Rd., Evanston, IL 60208 USA. Email: mamakos@u.northwestern.edu}

\setstretch{2} %

\section{Introduction}

In the evidence-based practice (EBP) movement, randomized trials are prioritized since, by design, they provide an unbiased estimate of the average treatment effect of an intervention \parencite{guyatt1993,chassin1998,imbens2010,sanderson2002,davies1999}. As the EBP movement has extended --- from medicine into policy, education, and social welfare ---  it has necessitated the development of new methods for improving randomized trials, including new study designs (e.g.,  \cite{hemming_stepped_2015, murphy_experimental_2005, collins_multiphase_2007}); methods for adjusting for biases resulting from attrition, noncompliance, and measurement error (e.g., \cite{frolich2014, bloom_accounting_1984}); and methods for improving statistical power (e.g., \cite{raudenbush2000, hussey2007}).

Adequate statistical power is now routinely required in both grant proposals and publications and a variety of tutorials, workshops, and software have been developed to help support this goal (e.g., \cite{spybrook_optimal_2011, dong_power_2018}).  These approaches to statistical power typically involve disaggregating the statistical power or the minimum detectable effect size into “design parameters” (e.g., \cite{cohen_statistical_2013,lipsey_design_1990, bloom_minimum_1995}). In cluster randomized trials, for example, these parameters include: the expected effect size ($\delta$), the number of clusters ($m$), the within cluster sample size ($n$), the intraclass correlation ($\rho$), the proportion of between-site variation that can be explained by covariates ($R^2$), and the proportion of clusters in treatment ($\pi$) \parencite{raudenbush_statistical_1997}. Software allows researchers to examine how different design parameters affect power, thus providing insights into how to better design their studies. For example, these indicate that if sample sizes are fixed, in order to increase statistical power, one might choose a design with equal allocation ($\pi = 1/2$), a sample that is fairly homogeneous ($\rho < .05$), and an outcome measure that is well aligned to the intervention, thus resulting in a large effect size ($\delta > 0.5$). 

But the questions EBP asks are broader than that of isolating and testing hypotheses about the average treatment effect. Indeed, EBP asks not “what is the average causal impact of this intervention in this study?” but instead, “what will the effect of this intervention be in [insert setting] or for [insert type of person]?” In the medical community, this is often framed as the need for individual treatment effects, as in precision medicine \parencite{hodson2016}. In the social sciences, the heterogeneity revolution has called into question the stability of causal effects outside of the confines of typical research environments \parencite{bryan2021}. And in education research, we see this in questions regarding “what works, for whom, and under what conditions”, with a focus on helping schools decide which interventions might work for them and their students. In each of these cases, the question is certainly one of causality, but the focus is not on the past, but instead on the future. Similarly, this question asks not about a single average effect, but about unit-specific effects (plural). 

Over the past decade, three streams of methods developments have buttressed this interest in conditional and unit-specific treatment effects. The first stream has focused on methods for causal prediction. This includes parametric approaches (including regression) and non-parametric, machine learning methods, including random forests, Bayesian additive regression trees (BART), and causal forests (e.g., \cite{athey2015, marshall2018, hahn_bayesian_2020}). For example, in 2018, the American Causal Inference Conference’s annual data challenge pitted these methods against one another Kaggle-style (e.g., \cite{hahn_atlantic_2019}). The second stream has focused on testing hypotheses regarding sources of treatment effect heterogeneity. Here there are questions regarding how to identify true moderators (as opposed to spurious associations; see \cite{vanderweele_explanation_2015}), as well as methods for improving the power of these tests (e.g., \cite{spybrook_progress_2016, dong_power_2018, tipton_beyond_2021}).  The third stream aggregates across this heterogeneity, focusing instead on how to generalize or transport average treatment effects from randomized trials to different populations (e.g., \cite{tipton_generalizability_2023, tipton_improving_2013, hartman_sample_2015, stuart_use_2011}). Developments here have focused on how eligibility and sample selection bias can affect both average causal effects and subgroup effects, as well as approaches for reducing this bias. 

But as these methods become increasingly integrated into practice --- thus meeting the promise of EBP --- there emerges a disconnect between what randomized trials are designed to do and how they are being used. That is, existing requirements and methods ensure that studies are designed to have adequate power for tests of average treatment effects, yet the data from such trials are being used to predict unit-specific treatment effects. This disconnect provides the motivation for this paper. Here we ask: How would we design randomized trials if, from the beginning, our goal was to predict unit-specific causal effects? In this framing, the goal of EBP is prediction, not hypothesis testing, and as such our focus is not on maximizing statistical power but on minimizing prediction error. To make progress on this we focus here on a two-group simple randomized experiment and predictions based upon parametric models estimated using OLS regression. We do so since this affords closed form expressions that parallel those found in the broader design and power analysis literature for randomized trials. 

%To preview our findings, we focus our paper around the goal of minimizing the mean squared prediction error (MSPE) across a defined population of units in need of predicted treatment effects. We show that this MSPE is a function of the standard error of the average treatment effect, as well as additional components that have to do with the number of covariates included and the degree of residual treatment effect variation. As a result, in small samples, unless strong moderators are available, the average treatment effect may have smaller prediction error than models including moderators. Furthermore, when the sample is not randomly selected from the target population --- what is referred to as 'covariate shift' or 'distribution shift' in the prediction literature --- we show that prediction errors (MSPE) can be substantially larger. 

The paper proceeds as follows. In Section \ref{sec:ate}, we provide an overview of current methods for the design and analysis of RCTs. In Section \ref{sec: pred_of_ute}, we introduce the the problem of prediction, focused on predicting unit-specific treatment effects using a parametric model, and deriving formulas for measuring the accuracy of these predictions. In Section \ref{sec: pred_and_gen}, we extend this to the (common) situation in which the estimation (i.e., source, training) sample is not drawn from the prediction (i.e.,target, test) population for whom predictions are desired. In Section \ref{sec: example}, we provide an example illustrating our findings, with a focus on small RCTs. We then conclude the paper with a discussion of the implications for planning studies and for predicting causal effects from RCTs.

%\begin{figure}[H]
%\vspace*{0.2cm}
%\centering
%\includegraphics[width=14cm]{figures/summary-table.png}
%\caption{\textcolor{blue}{Transferring this table from the docx notes for reference.}}
%\label{fig:summary-table}
%\end{figure}

\section{Average Treatment Effects} \label{sec:ate}

We begin by reviewing the literature on power analysis and generalizability, both of which will be central to the focus of this paper. To do so, we focus on the simplest study design --- the simple random control trial (RCT) --- in which $N$ units are randomized to a control ($n_0$) or treatment ($n_1$) condition respectively. We assume that for every unit $i = 1,...,N$ in the study, there are two potential outcomes, $Y_i(0)$ and $Y_i(1)$ and that, as a result of the Fundamental Problem of Causal Inference (Holland, 1986), we only observe one of these for each unit, i.e., we observe $Y_i = Y_i(0)(1-T_i)+Y_i(1)T_i$ where $T_i$ indicates if unit $i$ was randomly assigned to the treatment condition. 

\subsection{Designing for Sensitivity}

Using the observed data, we can estimate the average treatment effect using,
\begin{equation} \label{eq:1}
    \hat{\Delta} = \bar{Y}_1 - \bar{Y}_0
\end{equation}
where $\bar{Y}_1$ and $\bar{Y}_0$ are the sample means for those assigned to the treatment ($T=1$) and comparison ($T=0$) conditions respectively. It is easily shown that $\hat{\Delta}$ is an unbiased estimate of the sample average treatment effect (SATE). The standard error of the SATE can be estimated using \parencite{gerber_field_2008},
\begin{equation} \label{eq:2}
    \hat{SE}^2(\hat{\Delta}) = \frac{s_1^2}{n_1} + \frac{s_0^2}{n_0} 
\end{equation}
where $s_1^2$ and $s_0^2$ are estimates of the residual variances in the treatment and comparsion groups respectively. Notice here that we do not require that the true variances are equal (i.e., $\sigma_0^2 = \sigma_1^2$), though often in the literature on power analysis this is assumed. A test of the null hypothesis that $\Delta = 0$ can be conducted based upon the statistic
\begin{equation} \label{eq:3}
    t = \frac{\hat{\Delta}}{\hat{SE}(\hat{\Delta})}
\end{equation}
where under the null hypothesis, $t$ follows a t-distribution with degrees of freedom that can be estimated using a Satterthwaite approximation (since the variances differ).  

When designing an RCT, an important consideration is if the study design -- sample size, randomization process, etc -- will have enough sensitivity when estimating the average treatment effect. Design sensitivity can be thought of in terms of standard errors, statistical power, or the minimum detectable effect size (MDES).  The development of formulas and rules of thumb in all three cases typically involve simplifying assumptions. For example, it is common to assume that the residuals are normally distributed and share a common variance, $\sigma^2 = \sigma_1^2 = \sigma_0^2$. These simplifications allow for closed form expressions that convey the relationship between sensitivity and design parameters. For example, in the simple RCT, the MDES --- the smallest possible true effect size that could be detected with $1-\beta$ power and Type I error of $\alpha$ \parencite{bloom_minimum_1995}--- can be shown to be \parencite{dong_powerup!:_2013},
\begin{equation} \label{eq:mdes}
    MDES = M_{N -p - 2} \sqrt{\frac{1-R_p^2}{N\pi(1-\pi)}}
\end{equation}
where $N = n_0 + n_1$ is the total sample size, $\pi = n_1/N$ is the proportion in treatment, $R_p^2$ is the proportion of the within group variation that is explained by $p$ covariates included in the pooled model, and $M_{df} = t_{\alpha/2}(df) + t_{1-\beta}(df)$ is a multiplier. Notice here that the effect size (and thus MDES) is standardized in relation to the residual variation, $\Delta_s = \frac{\mu_1 - \mu_0}{\sigma}$. In more complex designs --- e.g., cluster randomized, multisite trials --- there are additional design parameters included in the MDES, such as the intraclass correlation ($\rho$), number of sites ($m$), and so on \parencite{raudenbush_statistical_1997}. 

These formulas for sensitivity can be solved for different parameters. For example, one might have a potential sample in mind and thus may be want to solve for $N$ to understand how many units need to be recruited. The formulas provide insights as well, regarding which design considerations are most consequential. In the standard error and MDES formulas above, for example, it is clear that the degree of residual variation ($\sigma$) is consequential. For example, a large degree of residual variation increases both the standard errors and the MDES and reduces the statistical power of the associated hypothesis test. In practice this means that researchers often favor more homogeneous samples (small $\sigma^2$).  When that itself is not possible, these formulas suggest that including covariates can improve sensitivity --- though keeping in mind that there is a push and pull here, with more covariates leading to greater $R_p^2$ (increased sensitivity) while also reducing the degrees of freedom (reduced sensitivity). 

\subsection{Designing for Generalizability}

In standard texts and methods related to the design of RCTs, the focus is nearly always on issues of sensitivity. This is because it is assumed --- explicitly or implicitly --- that the sample of $N$ units is itself the focus of the study, the sample can be conceived of as a random sample from some population, or that the treatment effect is fairly constant. More recently, this focus on internal validity to the exclusion of external validity has been called into question. \cite{imai_misunderstandings_2008} showed that if we are interested in the ATE for a population $P$ (i.e., PATE) --- not just the ATE in our sample (i.e., SATE) --- that,
\begin{align}
    bias(SATE) = E(SATE) - PATE = \Delta_{sample} + \Delta_{treatment} 
\end{align}
where $\Delta_{treatment}$ is bias resulting from either non-random assignment (or post-assignment attrition) and $\Delta_{sample}$ is bias resulting from non-random selection of the sample. They show that the ideal for causal generalization is a study with both random sampling and random assignment --- a design that has been exceedingly rare in practice. Studies in a variety of fields have followed, indicating that the samples involved in RCTs are typically not representative of target populations that are likely of interest for policy (e.g., \cite{stuart_characteristics_2017, tipton_toward_2020}). 

Here it is helpful to understand how this bias arises. Let us return to our potential outcomes framework, where now we add in the role of covariates. Here we include a set of $p$ covariates that potentially moderate the treatment effect; put another way, the relationship between each of these covariates and the observed outcome differs for those in treatment versus the comparison condition. Let $\bm{x}_i$ be a vector with elements $x_{ik}$, for $k = 1,...,p$ covariates. It is helpful here to standardize these covariates in relation to the sample $S$ so that $x_{ik|S} = \frac{x_{ik}-\mu_{x_k|S}}{\sigma_{x_k|S}}$. Thus we have,
\begin{align} \label{eq:6}
    Y_i(0) &= \mu_0 + \bm{x}_{i|S}'\bm{\beta}_0 + \epsilon_{0i}
    \\
    Y_i(1) &= \mu_1 + \bm{x}_{i|S}'\bm{\beta}_1 + \epsilon_{1i}
\end{align}
where we assume $E(\epsilon_{0i} | \bm{x}_{i|S}) = E(\epsilon_{1i}| \bm{x}_{i|S}) = 0$. Notice here that because the covariates are centered around the sample mean, the SATE can be defined simply as $SATE =\mu_1 - \mu_0$. However, now we can define the PATE for $P$ as
\begin{align}
    PATE &= SATE + \Delta_{sample}  
    \\
    &=SATE + (\bm{\mu}_{x|S} - \bm{\mu}_{x|P})'(\bm{\beta}_1 - \bm{\beta}_0) 
\end{align}
where $\bm{\mu}_{x|S} = E_S(\bm{x}_i)$ and $\bm{\mu}_{x|P} = E_P(\bm{x}_i)$ are vectors of average moderator values in the sample and population respectively.

Thus the sample selection bias that results is a weighted average of the covariate specific standardized mean differences between the sample and population. Clearly, what we have written here assumes that all of the relevant covariates are included --- what is referred to as a ``sampling ignorability condition'' \parencite{tipton_improving_2013, hartman_sample_2015, stuart_use_2011}. (Importantly, the standard error of the SATE --- even when used to estimate the PATE --- is not biased, since here the sampling variation is appropriately quantified with respect to the data collection process.)

%The fact that this bias is driven by underlying differences in the distribution of the moderators in the sample and population leads to a variety of indices of this bias. Stuart and colleagues thus propose to evaluate if a SATE may be biased for a PATE by examining separate standardized mean differences for each covariate in something akin to a ``balance'' table found in observational studies. Another approach is to reduce the dimensionality by comparing the sample to the population via a propensity score model. For example, here a logistic regression model is developed to predict sample membership based upon these moderator values. Thus, the overall similarity between the sample and population can be summarized by comparing the two distributions of these sampling propensity scores (e.g., via a ``generalizability index'', see 

When there is bias, a variety of methods have been developed to reduce this bias, including the use of weights (inverse probability, entropy), stratification, and regression (see \cite{tipton_generalizability_2023} for an overview). The application of these methods in practice, however, is often hampered by problems of undercoverage \parencite{tipton_improving_2013} --- parts of the population that are not represented at all in the sample; in related literature this is referred to as a violation of the common support or positivity assumption. When there is undercoverage, it is not possible to estimate the PATE without bias, and thus generalization to a smaller subset of the population may be the best that is possible. But even when all parts of the population are represented, if the sample $S$ is very different from population $P$, these adjustments tend to result in larger standard errors. That is, the standard error of the adjusted (unbiased) estimate of the PATE may be larger --- and significantly so --- than the standard error of the unadjusted (biased) SATE estimate. In practice, however, little is said of this bias-variance trade-off, since the focus of RCTs --- like in most causal studies --- is strongly on reducing bias.

The results from these adjustment methods suggest that a better approach, when feasible, is to design the study with one (or more) populations in mind and then to sample to represent this population. \textcite{tipton_stratified_2014} proposed using k-means cluster analysis to stratify on many possible moderators (since, in advance, which actually moderate is unknown). Within these strata, different selection methods are possible, including random and model-based approaches \parencite{litwok_selecting_2022}. When the focus is not only on estimating the average treatment effect but also on testing hypotheses regarding moderators of effects, \textcite{tipton_beyond_2021} shows that additional considerations for sampling are needed so that the sample has sufficient variation in the moderators to be tested. 

Finally, it is worth noting that this design approach fixes different parameters than the typical power analysis approach. By fixing the target population, it becomes clear that now the residual variation in treatment effects ($\sigma^2$) is fixed. As a result, it makes little sense to choose a more homogeneous sample if the goal is to explicitly generalize the ATE to a more heterogeneous population. In more complex designs, this means that related values --- like the intraclass correlation --- are also fixed. In practice, this means that larger sample sizes may be necessary and that finding and adjusting for covariates is even more important.

\section{Prediction of Unit-Specific Treatment Effects} \label{sec: pred_of_ute}

Underlying the generalizability concern --- that the sample and population ATEs differ --- is the assumption that treatment effects vary across units. But if treatment effects vary, why is the ATE of interest at all? This question is particularly salient for the EBP field, which provides estimates of ATEs in clearinghouses, encouraging decision-makers to transport these effects from the sample they were estimated on to a perhaps entirely different population. 

If treatment effects vary, our focus shifts from understanding how effective an intervention might be \emph{on average} to its effectiveness for a \emph{particular unit}. This unit might be an individual --- e.g., a student --- or an aggregate --- e.g., a school. This means our goal is one of predicting unit-specific impacts. Recently, there has been considerable development in methods for achieving this goal (for an overview, see \cite{hahn_atlantic_2019}). For example, Bayesian causal forests --- a version of Bayesian Additive Regression Trees (BART) --- have been shown to perform well (e.g., \cite{hahn_bayesian_2020}). In this paper, however, we are focused on design. Our question is: Under what conditions is prediction possible? Where are problems likely to arise? And if this is indeed our goal, how should studies be designed for this purpose? 

To answer these, in Section \ref{sec: sub_spec_of_model}, mirroring the literature on design, we narrow our focus to parametric linear models, which offer closed form expressions. Here we focus on a model that includes all moderators. In Section \ref{sec: sub_pred_error}, we derive measures of error relevant for prediction. Here we focus on the development of predictions for a population based upon an RCT conducted in a random sample from this population. In Section \ref{sec: sub_mod_selection}, we examine models that include only a subset of the moderators, with a focus on comparing the effect of additional moderators on error. Throughout, our focus is on deriving both formulas that can be useful when planning studies, as well as provide general insights regarding the importance of different parameters.

\subsection{Specification of the model} \label{sec: sub_spec_of_model}

To begin, assume we have a sample $S$ of $i = 1,...,N$ units, where $S$ is a random sample of some population $P_A$. For each unit $i$ recall that we have defined $Y_i(0)$ as the potential outcome of unit $i$ if this unit is assigned to condition $T = 0$, and $Y_i(1)$ is the potential outcome of unit $i$ if it is assigned to condition $T = 1$. Now, we also have available $k = 1,..,p$ covariates that moderate the treatment effect. For simplicity, we center the $k = 1,...,p$ covariates $x_{ik}$ around the mean in population $P_A$, $\mu_{x_k|A}$, and standardize them in relation to the population standard deviation, $\sigma_{x_k|A}$. We denote these standardized covariates using $x_{ik|A}$. Here $\bm{\beta}_0$ and $\bm{\beta}_1$ are $p$-dimensional vectors that relate the covariates to potential outcomes. The terms $\epsilon_{i0}$ and $\epsilon_{i1}$ are residual errors, with $E[\epsilon_{i0} | \bm{x}_{i|A}] = E[\epsilon_{i1} | \bm{x}_{i|A}] = 0$, $V(\epsilon_{i0}|\bm{x}_{i|A}) = \sigma^2_{0|\bm{x}}$ and $V(\epsilon_{i1}|\bm{x}_{i|A}) = \sigma^2_{1|\bm{x}}$. 

Using this notation, we can define the individual treatment effect $\delta_i$ for unit $i$ as,
\begin{align}
    \delta_i &= Y_i(1) - Y_i(0)
    \\
    &= (\mu_{1|A} - \mu_{0|A}) + \bm{x}_{i|A}'(\bm{\beta}_1 - \bm{\beta}_0) + (\epsilon_{i1} - \epsilon_{i0})
    \\
    &= \left[ \Delta_A + \bm{x}_{i|A}'\bm{\delta} \right] + \eta_i
\end{align}
Notice here that because of the standardization of the covariates, $E_A(\delta_i) = \Delta_A$ is the ATE in the sample and, because $S$ is a random sample of $P_A$, it is also the ATE for population $P_A$. The elements of the vector $\bm{\delta}$ correspond to covariates that moderate the treatment effect. Finally, notice that the part of this final equation in $[.]$ corresponds to the part of the unit specific treatment effect that is systematic and can thus be predicted, whereas the $\eta_i$ is the part that is idiosyncratic and cannot be predicted (\cite{ding_decomposing_2018}). Moving forward, we will assume that $E_A[\eta_i | \bm{x}_{iA}] = 0$ and that,
\begin{align}
    \tau_{A|x}^2 &= V_A(\eta_i | \bm{x}_i) 
    \\
    &= V_A(\epsilon_{i1} - \epsilon_{i0})  
    \\
    &= \sigma_{1|\bm{x}}^2 + \sigma_{0|\bm{x}}^2 - 2\rho_{01|\bm{x}}\sigma_{1|\bm{x}}\sigma_{0|\bm{x}}
\end{align}

Here, the correlation $\rho_{01|\bm{x}}$ between the residualized potential outcomes is unknowable because of the Fundamental Problem of Causal Inference \parencite{holland1986}. The fact that it is unknowable means that it is impossible to directly identify $\tau_{A|x}^2$. Instead, various approaches for bounding and sensitivity have been proposed (e.g., \cite{abadie_estimation_2008, fan_partial_2009, fan_sharp_2010}).

%\textbf{Correlation between potential outcomes.} While it impossible to identify this idiosyncratic variation in treatment effects, we can bound its value. If $\rho_{01|x} = 1$ then $\tau_{A|x}^2 = (\sigma_{1|\bm{x}} - \sigma_{0|\bm{x}})^2 $. Notice here that if $\sigma_{1|\bm{x}} = \sigma_{0|\bm{x}}$ then there is  no idiosyncratic variation. This corresponds to a case in which all the variation in the treatment effects is explainable --- and thus that any residual is left only because the model excluded covariates related to the outcome that did not moderate the treatment. Alternatively, if $\rho_{01|x} = 0$ then $\tau_{A|x}^2 = \sigma_{1|x}^2 + \sigma_{0|x}^2$.  

%An alternative assumption that could be reasonable is that the idiosyncratic variation in treatment effects is smaller than the variation in outcomes. In this case, we have 
%\begin{align*}
%    \sigma_{0|x}^2 &> (\sigma_{1|x} - \sigma_{0|x})^2 + 2(1 - \rho_{01|x}) \sigma_{1|x}\sigma_{0|x} 
%    \\
%    \rho_{01|x} &> \frac{1}{2} \left( \frac{\sigma_{1|x}}{\sigma_{0|x}} \right)
%\end{align*}
%Thus if the residual variances are equal, this implies that $\rho_{01|x} > 0.5$. Finally, this variance could be estimated based upon data, via a matching process (see Abadie CITE).  For now we remain agnostic about which assumption is realistic, though we will return to these and other approaches to bounding $\rho_{01|x}$ in later sections. 

\subsection{Prediction and Error} \label{sec: sub_pred_error}

We now assume that the purpose of our RCT is to build a model to predict $\delta_i$ for any unit $i$ in population $P_A$ with a vector of $p$ covariates $\bm{x}_i$. To do so, we will use OLS regression,  which provides closed form solutions that allow insights necessary for designing studies. In these models, we continue to standardize each of the $p$ covariates $\bm{x}_i$ with respect to the mean and standard deviation of population $P_A$; thus, we use the standardized vector $\bm{x}_{i|A}$ for prediction.

In order to predict the treatment effect for unit $i$ we need to predict each of the potential outcomes. For this, we can use the $n_0$ and $n_1$ units in the sample to build separate predictive models, resulting in the equations,
\begin{align}
    \ \hat{Y}_i(0) &= \hat{\mu}_{0|A} + \hat{\bm{\beta}}_{0}' \bm{x}_{i|A}
    \\
    \ \hat{Y}_i(1) &= \hat{\mu}_{1|A} + \hat{\bm{\beta}}_{1}' \bm{x}_{i|A}
\end{align}
From these $\hat{Y}_i(0)$ and $\hat{Y}_i(1)$, the predicted treatment effect for unit $i$ is,
\begin{align}
    \hat{\delta}_i &= \hat{Y}_i(1) - \hat{Y}_i(0)
    \\
    &= \hat{\delta}_A + ( \hat{\bm{\beta}}_{1} - \hat{\bm{\beta}}_{0} )' \bm{x}_{i|A}     
    \\
    &= \hat{\delta}_A + \hat{\bm{\delta}}' \bm{x}_{i|A} 
\end{align}
A question is thus how close this predicted effect is to the true treatment effect for unit $i$,
\begin{equation}
    \hat{\delta}_i -  \delta_i = (\hat{\Delta}_A - \Delta_A) + (\hat{\bm{\delta}} - \bm{\delta} )' \bm{x}_{i|A}  +  \eta_i
\end{equation}
In general, it is desired to have the difference $\hat{\delta}_i - \delta_i$ be as close to zero as possible, as this would indicate an accurate prediction of the treatment effect for unit $i$. 

\subsubsection{Measure of error for a specific unit} 

We need a measure of loss that provides a sense of the precision of this prediction for a specific unit. A common loss function is the squared prediction error (SPE), defined as
\begin{align}
    SPE(\hat{\delta_i}) &= E(\hat{\delta}_i - \delta_i \ | \bm{x}_{i|A} )^2 
    \\
    &= V(\hat{\delta}_A) + \bm{x}_{i|A}'V(\hat{\bm{\delta}}) \bm{x}_{i|A} + \tau_{A |\bm{x}}^2
\end{align}
This SPE is distinct for each unit $i$ since it depends upon the vector of covariates $\bm{x}_{i|A}$. The fact that the covariates are centered renders the ATE estimate independent of the moderator coefficient estimates. Thus, the second equality involves two terms that have to do with how well the average treatment effect (a function of the intercepts) and differences in slopes are estimated in the sample of $N$ units, while the third term has to do with the additional idiosyncratic treatment effect variation that is unexplained by the model. 

To further simplify the form of the SPE, we continue to assume that the sample of $N$ units is randomly assigned to a treatment and to a comparison condition and that this sample is randomly drawn from population $P_A$. Under this assumption it can be shown that,
\begin{align}
    SPE(\hat{\delta_i}) &= \left( \frac{\sigma_{0|\bm{x}}^2}{n_0}+ \frac{\sigma_{1|\bm{x}}^2}{n_1} \right) \left(1 + \bm{x}_{i|A}'\bm{\Sigma}_{\bm{x}|A}^{-1}\bm{x}_{i|A} \right) +  \tau_{A |\bm{x}}^2
\end{align}
where $\sigma_{0|x}^2$ and $\sigma_{1|x}^2$ are the residual variations in the two potential outcome prediction models, $\bm{\Sigma}_{x|A}$ is the variance covariance matrix of the standardized covariates in $P_A$, and $\tau_{A |x}^2$ is the total idiosyncratic variation in treatment impacts in $P_A$. For an observed unit $i$ in $P_A$, we can estimate this SPE using,
\begin{align}
    \hat{SPE}(\hat{\delta_i})  &= \left( \frac{s_{0|\bm{x}}^2}{n_0}+ \frac{s_{1|\bm{x}}^2}{n_1} \right) \left(1 + \bm{x}_{i|A}'\bm{S}_{x|A}^{-1}\bm{x}_{i|A} \right) +  \left[ (s_{0|\bm{x}} - s_{1|\bm{x}})^2 + 2s_{0|\bm{x}}s_{1|\bm{x}}(1-\rho_{01|\bm{x}}) \right]
\end{align}
where the values $s_{k|\bm{x}}^2$ are sample variances estimated in each of the two groups $k \in \{0,1\}$. Remember that this remains a function of $\rho_{01|x}$, which is unknowable. In practice, this means that a sensitivity approach may be required. Regardless, this SPE can be used to provide prediction intervals that convey the accuracy of the unit specific treatment effects. If we assume that the residuals are normally distributed, these can be created using critical values from the normal distribution ($z_{\alpha/2}$) and
\begin{align}
    (\hat{\delta_j} - z_{\alpha/2}\sqrt{\hat{SPE}}, \hat{\delta_j} + z_{\alpha/2}\sqrt{\hat{SPE}}).
\end{align}
In some cases such prediction intervals might be quite wide, indicating that while prediction is possible, it is not particularly informative. We will return to this topic in later sections.

\subsubsection{Combined measure of error for comparing and planning} 

Clearly, SPE varies across units in $P_A$. For model comparison and for planning purposes, it is therefore helpful to have an aggregate measure of the prediction error for the whole population $P_A$ in need of predicted treatment effects. A natural loss function to use here is the mean squared prediction error (MSPE), which averages the SPE across all units $i$ in population $P_A$ that need predictions,
\begin{align}
    MSPE(\hat{\delta_i}) &= E_{A}\left[ SPE(\hat{\delta}_i) \right] = E_A \left[ E(\hat{\delta}_i - \delta_i | \bm{x}_i)^2\right]     
\end{align}
Here we use the notation $E_{A}$ to indicate that this average is across all units in population $P_A$, which is our focus. The MSPE is a commonly used measure for assessing and comparing predictive models. It can be related to other metrics of model fit, such as Mallow's Cp and the Akaike Information Criterion (AIC), when residuals are assumed to be normally distributed \parencite{neter_applied_1996, hastie_elements_2009}.

In the parametric case considered in this paper, the MSPE can be shown to be,
\begin{align}
    MSPE(\hat{\delta}_i) &= \left( \frac{\sigma_{0|\bm{x}}^2}{n_0}+ \frac{\sigma_{1|\bm{x}}^2}{n_1} \right) \big(1 + p \big) +  \tau_{A|\bm{x}}^2 
\end{align}
which is a function of the residual variation in each group, the number of covariates, and the degree of idiosyncratic variation in effects remaining. Notice that this result is a straightforward extension of results found in standard regression texts.

For planning purposes, it is helpful to rewrite this MSPE in different terms. To do so, we first define $\tau_{*}^2 =\tau_{A}^2 / \sigma_{0}^2$ as the total treatment effect variation standardized by the variance in $Y(0)$. By writing $Y_i(1) = \bm{x}_i'(\bm{\beta_0} + \bm{\delta}) + \epsilon_i + \eta_i$, we can define two $R^2$ terms. The first --- $R_{0p}^2$ --- is the proportion of the variation comparison group variation ($\sigma_0^2$) explained by the $p$ covariates, while the second --- $R_{\tau p}^2$ --- is the proportion in treatment effect variation ($\tau^2$) explained by the $p$ covariates. Defining $R_{-a}^2 = 1 - R_a^2$, we can rewrite the MSPE as

\begin{align} \label{eq:mspe_full}
    MSPE(\hat{\delta}_i) 
    &= \frac{2 \sigma_{0}^2(1+p)}{n} \left[ R_{-0p}^2 + \tau_{*}\rho_{0\eta|x}R_{-\tau p}R_{-0p} + \tau_{*}^2R_{-\tau p}^2 \left( \frac{1}{2} + \frac{n}{2(1+p)}  \right) \right]
\end{align}

A proof for this is provided in Appendix A. Notice here that $\rho_{0\eta} = Corr(\epsilon_{0i},\eta_i)$ is the correlation between unit specific $Y_i(0)$ and unit specific treatment effects $\delta_i$, after conditioning on the covariates. When this correlation is positive, it indicates a treatment that increases disparities (i.e., larger effects for those with larger $Y(0)$ values). Like $\rho_{01|x}$, however, $\rho_{0\eta}$ cannot be identified. 

Writing the MSPE this way reveals the trade-offs between the number of covariates $p$ (that are estimated in each of the regressions) and the degree to which these covariates reduce the residual variation, both in terms of outcomes $R_{0p}^2$ and in terms of treatment effect moderators $R_{\tau p}^2$. Clearly, the inclusion of a covariate can increase the MSPE (through the $p+1$ term) or decrease it (through the $R^2$ terms). Thus, the inclusion of covariates that explain outcomes but do not moderate the treatment effect reduces $R_{0p}^2$ but does not reduce $R_{\tau p}^2$. The degree to which this matters, however, depends upon how much relative variation in treatment effects $\tau_{*}^2$ there is overall. 

\subsection{Model selection and prediction} \label{sec: sub_mod_selection}

Until now, we have focused on the general form of the MSPE, for the \emph{saturated} model that includes all $p$ moderators. Other models are possible, however, including: models with $r < p$ moderators; models in which the effects of covariates are assumed to be the same in both groups (ANCOVA); and a model with no covariates or moderators at all. Here the choice of ``best'' predictive model might compare a variety of these models, searching for that with the lowest relative MSPE. 

Here we focus on one important subclass of models: those that assume there to be no moderators of the treatment effect. This includes both the simple average treatment effect estimator (e.g., $\hat{\Delta} = \bar{Y}_1 - \bar{Y}_0$) and those that include adjustments for covariates (i.e., ANCOVA). We do so because in EBP, (covariate adjusted) ATE estimates are often collected and reported in individual papers and in evidence clearinghouses for use in decision-making regarding the adoption of an intervention. In effect, this approach predicts every unit specific treatment effect with the ATE (i.e., $\hat{\delta_i} = \hat{\Delta}_A$). A question, then, is if there are conditions under which this ATE estimate may outperform one that provides unit-specific predictions (i.e., using moderators). 

\subsubsection{ANCOVA and Raw Means Models} 
In estimation of an ATE, covariates are often included not as a means of predicting treatment effects, but as a way of reducing residual error. This is often referred to as an ANCOVA adjustment, since it assumes that the effects of the covariates are the same in both groups. This is akin to estimating a single model containing both the treatment and comparison units with an additive treatment, 
\begin{equation}
Y_j = \beta_0 + \Delta_A T_i + \bm{x}_{i|A}' \bm{\beta} + \epsilon_i
\end{equation}
A benefit of this model is that it involves estimation of fewer parameters ($p+2$ versus $2(p+1)$), while also reducing variance. However, this model results in a homogeneous treatment effect --- every unit $i$ in population $P_A$ is provided the same predicted treatment effect, $\hat{\Delta}_A$. 

In order to gain insight regarding this model, we focus on the balanced design in which $n_0 = n_1 = n$. In this case, the residual variation is pooled across both the treatment and comparison groups, resulting in the residual variation $\sigma^2 = (\sigma_0^2 + \sigma_1^2)/2$. The inclusion of $p$ covariates reduces this variation by a factor of $R_{-}^2 = (1-R^2)$. We can thus write the MSPE of this model using our notation as,
\begin{align} \label{eq:mspe_ancova}
    MSPE(\hat{\delta}_j | ANCOVA) &= \frac{\sigma^2(2+p)R_{-}^2}{2n} + \tau_A^2 \\
    &= \frac{\sigma_0^2(2+p)}{2n} \left[ R_{-0}^2 + \rho_{0\eta|x}\tau_{*}R_{-0} + \tau_{*}^2 \left(\frac{1}{2} + \frac{2n}{2+p} \right) \right]
\end{align}
A proof for this is provided in Appendix A. The first equality is straightforward, including two components; the first accounts for error that results from estimating $2+p$ coefficients from $2n$ observations, while the second accounts for the true variation in treatment effects. Notice that the inclusion of covariates does not affect this latter variation, since the model assumes a constant effect for all units. The second equality factors out of this a common variance; recall $\tau_*^2 = \tau^2/\sigma_0^2$ is a scaled version of the treatment effect variation.

A special case is a model in which no covariates are included. This is the unadjusted sample ATE estimator.  In this case, $p = 0$ and $R_0^2 = 0$. This results in the MSPE,
\begin{align} \label{eq:mspe_raw}
    MSPE(\hat{\delta}_j | raw) &= \frac{\sigma^2}{n} + \tau^2 \\
    &= \frac{\sigma_0^2}{n} \left[ 1 + \rho_{0\eta}\tau_* + \tau_*^2 \left(\frac{2n+1}{2}\right)  \ \right]
\end{align}
Notice here that in the first equality, the first term of the MSPE (which has to do with estimation error) goes to zero as the sample size increases, while the second term (which is the true variation) does not. 

\subsubsection{Comparing these models}

Given these models, under what conditions might the ATE provide a \emph{better} prediction of unit-specific treatment effects than a model that includes moderators?  To do so, we continue with our simplifying assumptions. First, we assume that $n_0 = n_1 = n$, as in a balanced design. Second, we assume that we are comparing nested models in which the same set of $p$ covariates are included in both the ANCOVA and the moderator models. In the ANCOVA model, the $p$ covariates are estimated in a single model that includes all $2n$ observations; in this model, an estimate of the ATE $\Delta_A$ is used to predict every unit specific effect (i.e., $\hat{\delta_i} = \hat{\Delta}_A$). In the moderator model, instead the relationship between the $p$ covariates and the outcome are estimated separately in the treatment and comparison conditions, and then subtracted to develop a unit-specific predictive model. A question then is when the ANOVA model provides a more accurate prediction than one including moderators. 

To study this, let $\tau_{*}^2 = \tau^2 / \sigma_0^2$ be the standardized treatment effect variation. The MSPE for a model with $p$ moderators (Equation \ref{eq:mspe_full}) can be shown to be smaller than the MSPE for a constant treatment effect model that adjusts for $p$ covariates (Equation \ref{eq:mspe_raw}) when 

\begin{align} \label{eq:minR2}
R_{\tau p}^2 \geq 1 - 
\left( \frac{
-(1+p)\rho_{0\eta}R_{-0p} \pm \sqrt{(1+p)^2 \rho_{0\eta}^2 R_{-0p}^2 - (1+p+n)\left[2(1+p)R_{-0p}^2 - nMSPE_p \right]}
}
{\tau_{*}(1+p+n)}
\right)^2
\end{align}

A proof for this is provided in Appendix A. As this equation shows, our preference for a model that includes moderators depends upon five parameters: the number of moderators $p$, the per group sample size $n$, the correlation $\rho_{0\eta}$ between comparison outcomes and treatment effects, the proportion of the comparison group variance explained by the covariates ($R_0^2$), and the degree of treatment effect heterogeneity $\tau_*^2$. In the next subsection we will investigate this empirically.

An important question is how this approach to model selection differs from one based on hypothesis testing. In both the hypothesis testing and prediction frameworks, nested models are compared (e.g., ANCOVA vs Moderator). However, the nature of the nested models differs. In the hypothesis testing approach, the null model is one in which there is no moderator relationship at all, i.e., the true $\tau^2 = 0$. In the prediction framework, we do not make this assumption. Instead, we assume that there may be variation in effects but we incorrectly model this using a simpler model. Thus, in the ANCOVA model, the values of the coefficients in $\beta$ are not the values under $Y(0)$, but instead the average values of those under $Y(0)$ and $Y(1)$. Similarly, the residual variation in the nested model is not simply a variance that is common to both $Y(0)$ and $Y(1)$ but instead is the average of these two. Put another way, here our question is which model better fits the data (in terms of minimizing mean squared error), instead of if an assumed model is true \parencite{breiman_statistical_2001, efron_prediction_2020}. 

\subsubsection{Simulation study}
In order to understand under what conditions we might prefer the ANCOVA model, we explore the minimum $R_{\tau p}^2$ required to prefer the alternative model. To do so, we examine the relationship between $R_{\tau p}^2$ and the parameters $n$, $p$, $\tau_*^2$, and $\rho_{0\eta}$. Importantly, while $n$ and $p$ are known to researchers, both $\tau_*^2$ and $\rho_{0\eta}$ are not. 

We focus on a range of treatment group sizes $n$ that are small to large, including values of $n$ between $10$ and $1000$. The small values are included since the units of interest may be groups (e.g., clusters, sites) and the studies may involve randomizing these groups; for example, we may desire treatment effect predictions for all sites (schools, hospitals) in a population based upon the results of a cluster-randomized trial. Typical cluster randomized trials can include as few as $n = 20$ sites in each treatment arm (assumed to be equal here). 

\begin{figure} 
    \centering
    %\hspace{-3cm}
    \includegraphics[scale=.75]{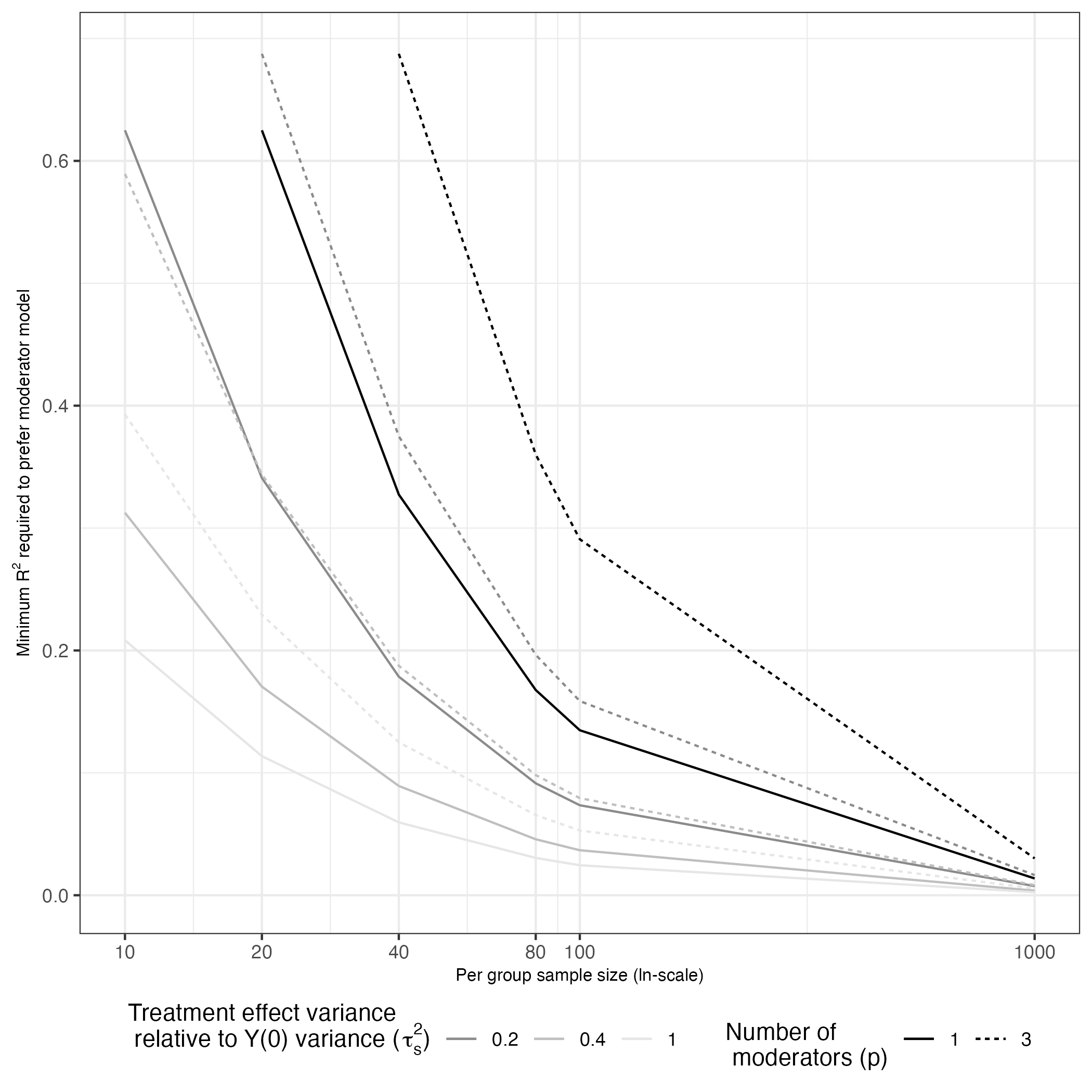}      
   \caption{Minimum required $R_\tau^2$ by sample size, degree of variation, and number of covariates. Values shown are for $R_0^2 = 0.5$ and $\rho_{0\eta} = 0$.}
  \label{fig:figpn}
\end{figure}

Examining Figure \ref{fig:figpn}, we can see that in studies with a small number of units, in order for moderator models to improve predictive accuracy, they need to explain significant portions of the variation in treatment effects, particularly when the overall degree of treatment effect variation is small or moderate. For example, with $n = 20$ units per arm, if the treatment effect variation is about 40 percent of the outcome variation ($\tau_*^2 = 0.40$), a model with $p = 1$ moderator would need to explain at least $100R_\tau^2 = 34$ percent of the variation in order to outperform an ANCOVA model. If one variable alone could not do this, a model with three variables would need to explain $68$ percent of the overall treatment effect variation. In comparison, in samples with $n = 100$ units per arm, these percents drop to $8$ percent and $16$ percent respectively.

Overall, this suggests that for models of unit-specific predictions of treatment effects to perform well, either substantially larger sample sizes are required or single moderators with large explanatory value need to be included. If treatment effects are not functions of only one or two variables, but instead vary in relation to a large variety of moderators (each with small effects), this means that very large samples are necessary to outperform the ANCOVA estimator. Put another way, in the small samples typical in cluster-randomized trials, the covariate-adjusted average treatment effect may offer the most accurate prediction of any unit's treatment effect, even if the true effects vary. 

\section{Prediction and Generalizability } \label{sec: pred_and_gen}

Until now, we have assumed that the sample of $N$ units in the RCT was a random sample from the population $P_A$ for whom treatment effect predictions are required. These are the same assumptions that are typical in the development of estimators of SPE for unit-specific predictions and MSPE for model comparisons. However, as the literature on methods for generalizing and transporting causal effects indicates, this assumption is likely not tenable in practice. That is, very often design decisions --- including the selection of the sample --- are made based upon a desire to improve sensitivity (reduce error) without regard to how the average treatment effect estimated in the sample will relate to ATEs for population of interest for policy and practice. More directly, researchers are often searching for and selecting the population $P_A$ that allows for a precisely estimated average treatment impact. In the case of prediction, this would be akin to minimizing the MSPE by purposely selecting a sample in which the treatment effect did \emph{not} vary, thus allowing for the ANCOVA or unadjusted estimators to dominate.

In this section, we take a different approach. In Section \ref{sec: predest}, we continue to assume that the sample is randomly drawn from a population $P_A$, but we now assume that unit-specific treatment effect predictions are desired for a different population $P_B$. We show that this shift in population results in less accurate predictions regarding unit specific treatment effects. In Section \ref{sec: adjust}, building on recent developments for predictive models under covariate shift, we show that the use of weighting adjustments can reduce the MSPE, but that these weighting methods come with a variance inflation penalty (relative to predictions in $P_A$). In Section \ref{sec: simstudy}, we provide examples of different population shifts to illustrate how large a penalty one might expect in practice, as well as general implications of these for designing studies. 

\subsection{Prediction and Error} \label{sec: predest}

To begin, we assume that in population $P_B$, there is a set of $r$ covariates $\mathbb{B}$ that moderate the treatment effect systematically. Similarly, we assume that in population $P_A$, there is a set of $p$ covariates $\mathbb{A}$ that moderate the treatment effect. Let $\mathbb{Q} = \mathbb{A} \cup \mathbb{B}$ be the union of these sets, resulting in $q$ covariates that moderate the treatment effect in one or both of these populations. Now, assume that we standardize each of these $q$ covariates with respect to population $P_A$; we do so since we will be estimating our model in $P_A$. For a unit $j$ in $P_B$ then we can define their true treatment effect as,
\begin{align}
    \delta_j = \Delta_A + \bm{\delta}_B' \bm{x}_{j|A} + \eta_j
\end{align}
Note that if there is a covariate here that moderates the effect in $P_A$ but not in $P_B$, it takes the value of zero in this vector $\bm{\delta}_B$. Now, we have available to us a treatment effect prediction model estimated on $P_A$, 
\begin{align}
    \hat{\delta}_j = \hat{\Delta}_A + \hat{\bm{\delta}}_A' \bm{x}_{j|A} 
\end{align}
Notice here that $\hat{\bm{\delta}}_A$ signifies that this estimate of the moderator coefficients is based upon the sample provided from $P_A$. 

\subsubsection{Bias}
Again, we are interested in understanding how close $\hat{\delta}_j$ is to $\delta_j$. Here we might begin by examining the bias,
\begin{align}
    bias(\hat{\delta}_j) = \bm{x}_{j|A}'\bm{\theta}_{B|A} + \eta_{j|B}
\end{align}
where $\bm{\theta}_{B|A} = (\bm{\delta}_A - \bm{\delta}_B)$ is the difference between the relationship between the standardized covariates $x_{jk}$ and treatment effects $\delta_j$ in population $P_A$ versus that in $P_B$.  One way to think of this is as ``extrapolation'' bias, which results from extrapolating a relationship beyond the support of $x_{jk}$. In the ML literature, this is referred to as the problem of ``distribution shift'', wherein the covariate distribution used in the training data (here $P_A$) differs from that in the test data (here $P_B$). 

Two examples help illustrate the source of this extrapolation bias. For the first, assume that the covariate results from a categorical variable that indicates if a unit $j$ is located in an urban, suburban, town, or rural area. Perhaps this is encoded as three dummy variables, with urban as the reference. Suppose that the true treatment effects in $P_B$ differ across these areas, and that all areas are represented in $P_B$. However, suppose that $P_A$ by design only included urban areas --- then this would mean that it was not possible to estimate differences between treatment impacts across these areas (in the generalization literature, this is referred to as ``undercoverage''; \cite{tipton_improving_2013}). In our framework, this would amount to setting the coefficient associated with the $k^\text{th}$ covariate to be zero in one population ($\delta_{kA} = 0$), while it is non-zero in the other ($\delta_{kB} \neq 0$). For the second example, consider a continuous covariate $x_k$ in which the support of the covariate differs in $P_B$ relative to $P_A$. For example, the support of $x_{k}$ in $P_B$ might be larger than in $P_A$. Suppose that within $P_A$ the relationship between $x_{jk}$ and $\delta_j$ is linear. It is easy to see that such a model could be correct in $P_A$ and yet incorrect in $P_B$ -- e.g., if in the larger range of $x_k$ values, the relationship was non-linear.

%Figure 2 illustrates this case.

%\begin{figure}[H]
%    \centering
%    %\hspace{-3cm}
%    \includegraphics[scale=0.24]{figures/extrapolation.jpg} 
%   \caption{Illustration of bias from extrapolation when moving from estimation in $P_A$ to prediction in $P_B$}
%  \label{fig:betfig}
%\end{figure}

\subsubsection{Unit specific error}
Again, we can quantify the error in our predictions using the squared prediction error (SPE), which is a function of both this bias and the sampling variance from the estimation of the model in $P_A$,
\begin{align}
   E\left[ (\hat{\delta}_j - \delta_j )^2 \ | \bm{x}_{j|A} \right] &= V \left( \hat{\Delta}_A + \hat{\bm{\delta}}_A' \bm{x}_{j|A} \ | \bm{x}_{j|A} \right) + \bm{\theta}_{B|A}' \bm{x}_{j|A}'\bm{x}_{j|A} \bm{\theta}_{B|A}  + \tau^2_{B|\bm{x}} \\
   &= \left( \frac{\sigma_{0|\bm{x},A}^2}{n_0}+ \frac{\sigma_{1|\bm{x},A}^2}{n_1} \right) \left( 1 + \bm{x}_{j|A}'\bm{\Sigma}_{A}^{-1} \bm{x}_{j|A} \right) + \bm{\theta}_{B|A}' \bm{x}_{j|A}'\bm{x}_{j|A} \bm{\theta}_{B|A}  
   + \tau^2_{B|\bm{x}}
\end{align}
Notice here that there are now three terms. As before, the first term has only to do with how well the coefficients are estimated in the sample of $N$ units in $P_A$. The second term is now the squared bias. Finally, the third term refers to the unexplained, idiosyncratic treatment effect variation in $P_B$. We can further expand this as,
\begin{equation}
\tau^2_{B|\bm{x}} = \sigma_{0|\bm{x},B}^2+\sigma_{1|\bm{x},B}^2 - 2\rho_{01|\bm{x},B}\sigma_{0|\bm{x},B}\sigma_{1|\bm{x},B}
\end{equation}
This is a function of the residual variances $\sigma_{0|\bm{x},B}^2$ and $\sigma_{1|\bm{x},B}^2$ and the correlation $\rho_{01|\bm{x},B}$. All three of these parameters have to do with relationships found in $P_B$ \emph{not} in $P_A$. If $P_B$ differs markedly from $P_A$ -- as occurs when $P_A$ was selected to be homogeneous --- it is not hard to imagine that residual variances from $P_A$ might not apply to $P_B$. Any estimation of this variance now requires further assumptions or data than in the $P_A$ case.
 
In general, this indicates that a problem that results when moving to $P_B$ is how such prediction error can be estimated for a unit $j$. Parameter values with subscripts $A$ can be directly estimated from the sample in $P_A$. But parameter values with subscripts involving $B$ are unknowable. For example, we cannot know the degree of bias in the treatment effect moderators $\bm{\theta}_{B|A}$, since such relationships can only be estimated in $P_A$. Similarly, we cannot know if the variation due to idiosyncratic variation in treatment effects $\tau_{B|\bm{x}}^2$ is the same as that in $A$; even decomposing this requires extant information on the degree of residual variation in the outcomes in $B$. Despite this, one may be tempted to estimate this error using the previously defined SPE formula,
\begin{align}
   \hat{SPE}(\hat{\delta}_j) &= \left( \frac{s_{0|\bm{x},A}^2}{n_0}+ \frac{s_{1|\bm{x},A}^2}{n_1} \right) \left( 1 + \bm{x}_{j|A}'\bm{S}_{A}^{-1} \bm{x}_{j|A} \right) + \left[ s_{0|\bm{x},B}^2+ s_{1|\bm{x},B}^2 - 2\rho_{01|\bm{x},B}s_{0|\bm{x},B}s_{1|\bm{x},B} \right]
   \\
   &= SPE(\hat{\delta}_j) - \left[ \bm{x}_{j|A}' \bm{\Theta}_{B|A} \bm{x}_{j|A}' + (\tau_A^2 - \tau_B^2) \right]
\end{align}
However, as the second line indicates, this estimate of the error can be biased when $P_A \neq P_B$. In general, this bias could be positive or negative, leading to over or under estimates. We will return to this with special cases later. 

\subsubsection{Average error across units}
It is helpful again to summarize this error across the whole of population $P_B$ for whom we seek predictions. In order to develop the MSPE in this case, we need to define a few more parameters. Recall that we are standardizing our $q$ covariates with respect to $P_A$; we do so since this is how the covariates were standardized in $P_A$, where our model was estimated. Here again keep in mind that we are not making any other assumptions regarding the distribution of covariates.  Now, in $P_B$ it can be shown that the MSPE can be written,
    \begin{equation} \label{eq: m_and_d}
    MSPE(\hat{\delta}_j | B) = \left( \frac{\sigma_{0|\bm{x},A}^2}{n_0}+ \frac{\sigma_{1|\bm{x},A}^2}{n_1} \right) \left(1 + D_{B|A} + M_{B|A} \right) + \mathrm{tr}[\bm{\Theta}_{B|A}\bm{\Sigma}_{B|A}] + \tau_{B|\bm{x}}^2
    \end{equation}
where $M_{B|A} = (\bm{\mu}_B - \bm{\mu}_A)\bm{\Sigma}_{A}^{-1}(\bm{\mu}_B - \bm{\mu}_A)$ is the squared Mahalanobis distance between populations $P_B$ and $P_A$ and $D_{B|A} = \mathrm{tr} \left[ \bm{\Sigma}_{A}^{-1} \bm{\Sigma}_{B} \right]$ is proportional to the Burg divergence. Equation \ref{eq: m_and_d} falls from properties of quadratic forms, where for a general random vector $\bm{z}$ and matrix $\bm{A}$, $E(\bm{z'Az}) = \bm{\mu_z'A\mu_z} + tr(\bm{A\Sigma_z})$. Readers might note that the quantity $D + M \equiv D_{B|A} + M_{B|A}$ is proportional to the Kullbach-Leibler divergence between two multivariate normal distributions (though we make no such normality assumptions here). 

In the special case when $p = 1$, the formula for $D + M$ can be shown to simplify to,
\begin{align}
D_{B|A} + M_{B|A} &= \frac{\sigma_B^2}{\sigma_A^2} + \left(\frac{\mu_B - \mu_A}{\sigma_A} \right)^2 
    \end{align}
Notice that this is a function combining differences in the first and second moments of the distributions of the moderators in $P_A$ and $P_B$. Examining this, we can see that when $P_A \equiv P_B$ , this function equals $1$. In general, when the two distributions differ, these are values are likely greater than $p$. However, examining this formula shows that this is not the smallest possible value. Notice that this can be further minimized by selecting the estimation population $P_A$ so that it is \emph{more heterogeneous} than $P_B$, i.e., so that $\sigma_A^2 >> \sigma_B^2$.  

Finally, note here that the first two terms (multiplied) in the MSPE can be estimated directly from data available in $P_A$ and $P_B$. However, like SPE, the latter two terms are more complex, as they require information that is directly unknowable --- the degree of bias in the moderator coefficient estimates $\hat{\bm{\delta}}$ and the degree to which the variation in residual variation in $P_B$ is more or less the same as in $P_A$.  

\subsubsection{Error when using sample ATE}
In the previous section, we showed that when $P_A \equiv P_B$, in small samples, the ATE has smaller MSPE than unit-specific treatment effects. For completeness, we investigate the MSPE here when using the ATE estimate from $P_A$ as the predicted unit-specific treatment effect for all units in $P_B$. More formally, let $ \hat{\delta}_j = \hat{\Delta}_A$ for all units $j = 1,...,N$ in population $P_B$. It is straightforward to show that the MSPE across all units in $P_B$ is,
\begin{align} \label{eq:mspe_pate}
    MSPE(\hat{\delta}_j | B ) &= \left( \frac{\sigma_{0|\bm{x},A}^2}{n_0} + \frac{\sigma_{1|\bm{x},A}^2}{n_1} \right) + 
    (\bm{\mu_B - \mu_A})' \bm{\Theta}_{B} (\bm{\mu_B - \mu_A})
    + \tau_{B}^2 \\
    &\leq  \left( \frac{\sigma_{0|\bm{x},A}^2}{n_0} + \frac{\sigma_{1|\bm{x},A}^2}{n_1} \right) + 
   \left(\bm{\delta}_B' \bm{\Sigma}_A \bm{\delta}_B \right) M_{B|A}
    + \tau_{B}^2
\end{align}
where $\bm{\Theta_B} = \bm{\delta}_B \bm{\delta}_B'$. Notice that when $M_{B|A} = 0$, Equations \ref{eq:mspe_ancova} and \ref{eq:mspe_pate} differ only in the idiosyncratic error term $\tau_B^2$. This residual error is specific to the prediction population. This result aligns with previous work on the generalization of average treatment effects, which has focused on the reduction of bias via efforts to weight sample data (from $P_A$) to have the same moderator means as that in the target ($P_B$) population \parencite{stuart_use_2011, tipton_improving_2013}.

\subsection{Possible adjustments} \label{sec: adjust}
As a result of the shifting covariate distributions between $P_A$ and $P_B$ we see that the accuracy of the model decreases. That is, the MSPE is now a function of the degree of shift, which has to do with both $M_{B|A}$ and $D_{B|A}$. This suggests that a better approach would be to take into account this covariate shift in the estimation process. To do so, first stack the data so that $P = P_A \cup P_B$ and let $Z_i$ indicate if a unit $i$ is in $P_A$.

For general prediction problems, \textcite{shimodaira_improving_2000} proposes the use of weighted regression, with weights
\begin{equation}
w_i = \frac{Pr(\bm{x}_i|Z_i = 0)}{Pr(\bm{x}_i| Z_i = 1)}.  
\end{equation}
Notice that these weights require knowledge of the joint distribution of the covariates; e.g., Shimodaira assumes that the covariates are normally distributed. \textcite{steingrimsson_transporting_2023} provide an alternative specification of weights that are simpler to specify. They show that by applying Bayes' rule, the \textcite{shimodaira_improving_2000} weights are proportional to the inverse odds that a unit is in $P_A$ (relative to $P_B$),
\begin{equation} \label{eq: inverseodds}
 w_i \propto \frac{Pr(Z_i=0|\bm{x}_i)}{Pr(Z_i = 2|\bm{x}_i)} = \frac{1 - Pr(Z_i = 0| \bm{x}_i) }{Pr(Z_i = 0| \bm{x}_i)} 
\end{equation}
This formulation suggests that weights can be estimated using logistic regression or a variety of other methods found in the propensity score literature (see \cite{stuart_matching_2010} for an overview). 

When predicting treatment effects, this means incorporating these weights into estimation of both $\bm{\beta}_1$ and $\bm{\beta}_0$. To do so, for each $i = 1,...,N$ units in $P_A$, define the weight $w_i$ as above.  For each of $k = 0, 1$, define a weight matrix $\bm{W}_k$ as an $N_k \times N_k$ diagonal weight matrix.  Using these weights, for $k = 0,1$ calculate
\begin{equation} 
\hat{\bm{\beta_k^w}} = (\bm{X_k'W_kX_k})^{-1}\bm{X_k'W_kY_k}. 
\end{equation}

For a unit $j$ in $P_B$, a treatment effect can thus be predicted using $\hat{\delta}_j^w = \bm{x}'_j (\bm{\beta_1^w} - \bm{\beta_0^w})$. 
Notice that this is the same form as before, but now weighted regression is used instead.

\textcite{steingrimsson_transporting_2023} shows that two assumptions are required for this weighting estimator to result in an unbiased estimator of both the unit-specific treatment effects and the degree of prediction error (MPSE):

\begin{itemize}
    \item A1: \emph{Conditional independence of the outcome Y and the population.} For every $\bm{x}$ with positive density in $P_B$, $f(X=x, Z=0)> 0$, $f(Y|\bm{X = x}, Z = 1) = f(Y|\bm{X = x}, Z = 0).$
  
    \item A2: \emph{Positivity.} For every $\bm{x}$ such that $f(X = x, Z = 0) \neq 0$, we have $Pr(Z = 1|X = x) > 0. $
\end{itemize}

Assumption A2 means that every covariate pattern found in the target population $P_B$ also exists in population $P_A$. This ensures that the predictions of $\delta_j$ for units in $P_B$ do not require extrapolations beyond the estimation data. Assumption A1 implies that for $k = 0,1$, the $\bm{\beta}_k$ estimated in $P_A$ can be transported to $P_B$ and that the estimate $\hat{\delta}_j^w$ of the treatment effect for unit $j$ is unbiased. Furthermore, Assumption A1 also implies that the estimator of the MSPE used based upon the sample data from population $P_A$ is unbiased for the estimator of the MSPE in population $P_B$. In an appendix, \textcite{steingrimsson_transporting_2023} provides an example illustrating a violation of this assumption. 

The use of weights comes at a cost in terms of sensitivity. Let $M_{VIF}$ be a multiplier that indicates the degree of variance inflation due to use of weighting, where
\begin{equation}
M_{VIF} = \frac{Var(weighted)}{Var(unweighted)}.
\end{equation}
Applying the weights and assumptions of \textcite{steingrimsson_transporting_2023} to each of the regressions separately and combining them provides, 
\begin{align}
    MSPE(\hat{\delta}_j^w) = M_{VIF} \left( \frac{\sigma_{0|\bm{x}}^2}{n_0}+ \frac{\sigma_{1|\bm{x}}^2}{n_1} \right) \big(1 + p \big)  + \tau_{A|\bm{x}}^2 
\end{align}
Notice here that as a result of Assumptions A1 and A2, the term $\bm{\Theta'\Sigma} = \bm{0}$ is not included; similarly, A1 implies that $\tau_B^2 = \tau_A^2$, which enables estimation of the idiosyncratic treatment effect variation in $P_B$ from data in $P_A$.

The resulting multiplier $M_{VIF}$ can be approximated using Kish's design effect \parencite{kish_weighting_1992}, which is a function of the coefficient of variation of the weights,
\begin{equation} \label{eq:vif}
M_{VIF} = 1 + \frac{V_A(w_i)}{E_A(w_i)^2}.
\end{equation}
Alternatively, this variance inflation can be written in terms of an effective sample size $N_e$. To do so,
\begin{equation}
N_e = N / M_{VIF}.
\end{equation}
This effective sample size indicates how much smaller a sample $N_e$ from $P_B$ could be to have the same precision as the sample of size $N$ from $P_A$. 

Importantly, since this $M_{VIF}$ multiplier does not require outcome data, it can also be used for planning purposes. To do so, researchers would begin by specifying a target population $P_B$ and several possible populations for estimation (samples) $P_A$. For each, they could then calculate the distance $M + B$, if the positivity assumption is met (by exploring the common support), and the expected variance inflation penalty. By conducting this analysis, a researcher may realize that a sample from some population $P_A$ is not adequate for making predictions for units in $P_B$. 

\subsection{Case study} \label{sec: simstudy}

In this section, we explore the size of this distribution shift penalty using a case study in education. The Common Core of Data provides a census of public schools in the United States. States are required to submit demographic data on schools in a common form; however, this data is not always complete (e.g., some states may not report certain variables).  For this study, we narrow this population to focus on elementary schools and select five potential moderators of a treatment effect: the number of schools in a school district (District Size); the number of students in a school (School Size); the Student to Teacher ratio (Stu/Tch); the proportion of students receiving free- or reduced-priced lunch (an indicator of poverty; PropFRL); and an indicator of if the school is located in an urban locale (Urban). We restrict ourselves to the population of schools that have no missing data on these five variables; this results in a population of $9,175$ schools in $P_B$.

\subsubsection{Covariate shift across populations}
\begin{figure}
    \centering
    %\hspace{-3cm}
    \includegraphics[scale=.75]{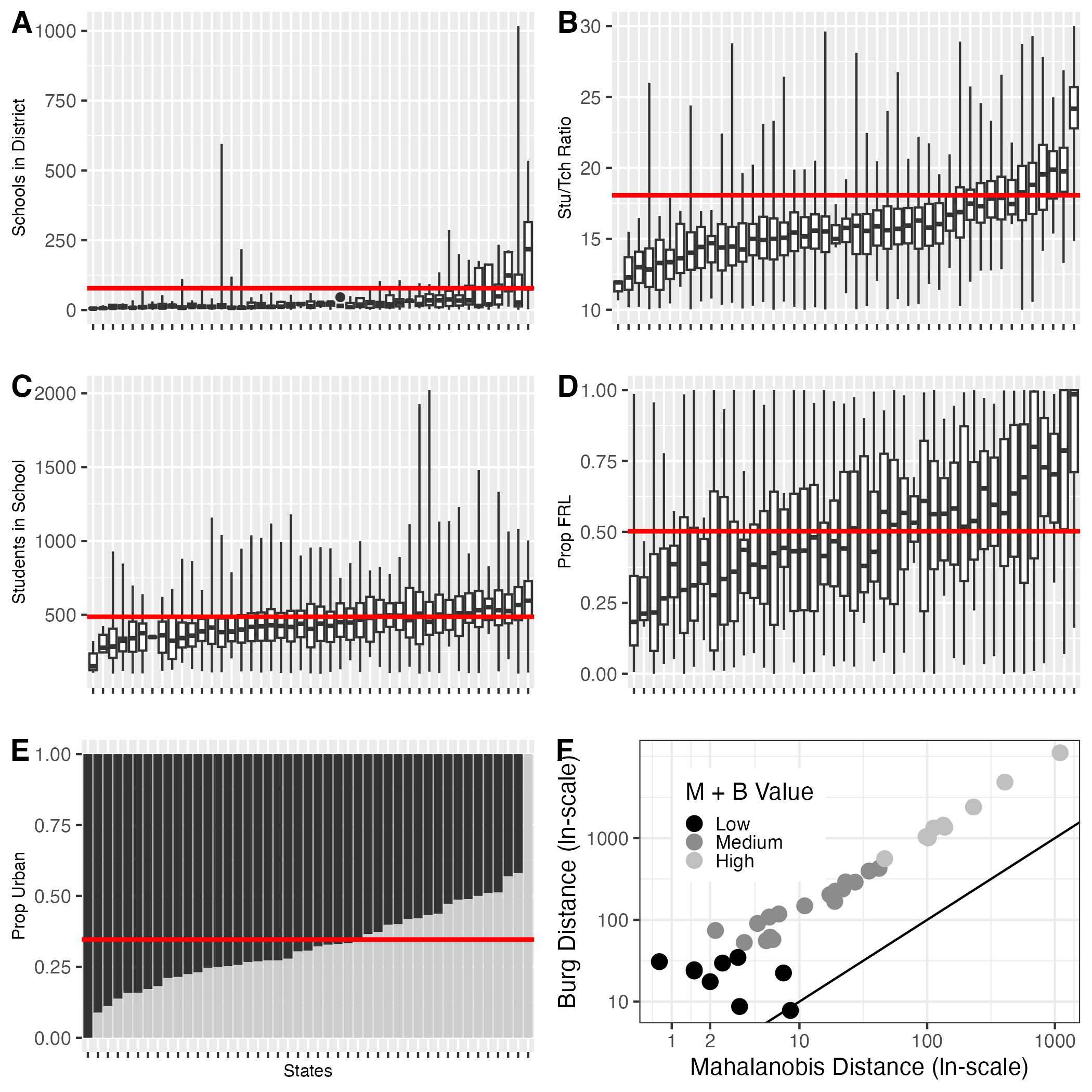}      
   \caption{A - E. Covariate distributions for state populations of elementary schools, ordered by median. F. Comparisons of mean and variance differences for state versus U.S. populations of elementary schools.)}
  \label{fig:states}
\end{figure}

We then envision a situation in which a study could take place in a single state ($P_A$) while the ultimate goal would be to predict treatment effects for all public elementary schools in the U.S. ($P_B$). Here we only include states with at least $40$ schools with complete data; the resulting $35$ states include populations of between $46 - 2,403$ schools, with a median of $141$ schools. In Panels A-E of Figure \ref{fig:states} we provide 5-number summaries via boxplots for each of these states across the five covariates. Each panel is ordered from smallest to largest with respect to the median values. The solid horizontal line indicates the average value in population $P_B$. Importantly, notice that the minimum and maximum values vary considerably across these state populations. For example, there are some states with only small school districts, while others include a range of school district sizes. These range differences will ultimately affect the positivity assumption (A2) required for making predictions in $P_B$.

We next calculated the Mahalanobis distance $M$ and the Burg distance $D$, as defined in Equation \ref{eq: m_and_d}. Recall that when the population $P_A \equiv P_B$ we would expect $M + D = p = 5$, and that values above $p = 5$ greatly increase the average prediction error (MSPE). Panel F, which is on a log-scale, indicates that these values are very high for most of these state populations, with combined values often above $100$ or even $1000$. Panel F also illustrates that the differences between $P_A$ and $P_B$ are not limited to mean differences (Mahalanobis distance), which has been the focus of much of the causal generalization literature. The Burg distance indicates that additionally, the variances and covariances in $P_A$ tend to be smaller than in $P_B$. In general, the Burg distance for these state populations tend to be an order of magnitude larger than the mean differences. 

\subsubsection{Weighting adjustments}

\begin{figure}
    \centering
    %\hspace{-3cm}
    \includegraphics[scale=.55]{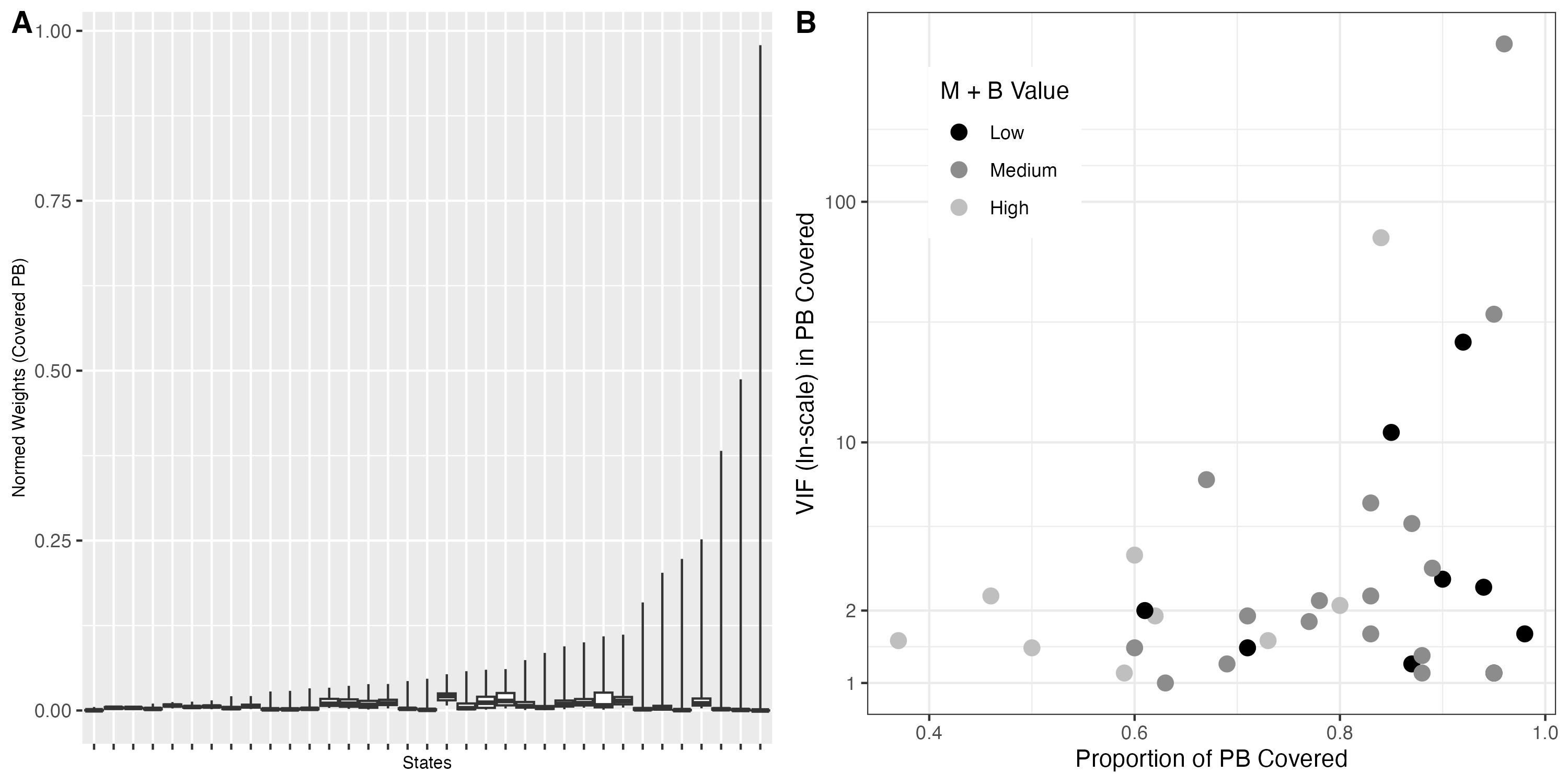}   
   \caption{A. Distributions of normed inverse odds weights for states (weights comparing state to U.S. population of elementary schools. B. Comparison of degree of coverage of the U.S. population by variance inflation and overall degree of difference.}
  \label{fig:vif}
\end{figure}

Given the large covariate distribution shifts, in order to predict treatment effects in $P_B$, weighting adjustments would be required. For each state population, we calculated the inverse odds weights, as defined in Equation \ref{eq: inverseodds}. In order to meet the positivity assumption, we examined the common support of the distribution of these weights and excluded U.S. schools ($P_B$) outside the support of $P_A$. This meant that treatment effect estimates would not be possible for some proportion of the target population, i.e.,'undercoverage'. The x-axis of Panel B of Figure \ref{fig:vif} shows the range of coverage of $P_B$ across these state populations. Notice, for example, that some states are so different from $P_B$ that it would be possible to predict treatment effects for less than 40-percent of the U.S. population of schools.

For each state population $P_A$, we then normed the weights so that they summed to one. The distribution of these weights is then provided in Figure \ref{fig:vif} Panel A. Notice that in fives states, one school carries 25-percent or more of the weight. Large weights of this sort are often trimmed in practice \parencite{lee_weight_2011}, but this trimming introduces bias; for this reason, we do not trim the weights here. The VIF was then calculated based on these weights using Equation \ref{eq:vif}. Panel B of Figure \ref{fig:vif} shows the relationship between the degree of coverage of the target population $P_B$ and the variance inflation that results from estimation in $P_A$. Here the shade of the data points indicates the relative size of the covariate distance $M + D$. Examining the y-axis, we see that for roughly half of the states, the VIFs are between 1 and 2. In these cases, the variance inflation is small, though not insubstantial. For example, a VIF of 1.5 indicates an approximate 50-percent increase in the prediction error, relative to estimating treatment effects in $P_B$ directly. In the other half of states, however, these VIFs are considerably larger - with 5 involving VIFs greater than 10. Further examination of the data indicates that these cases correspond to those with large weights (found in Panel A). 

Overall, this analysis indicates a few important trends. First, similar to when generalizing an ATE estimate from a sample to population, as mean covariate differences increase, so too do the variance penalties due to adjustment \parencite{tipton_implications_2017}. Second, when predicting unit-specific effects, these differences and penalties are also affected by differences in variances and covariances. When mean differences are large, these variance and covariance differences also tend to be large. Just like the generalization situation, the result is that there can be a large degree of undercoverage --- parts of the population for whom no treatment effect prediction is possible --- and/or increased variance.

\section{Example: Planning a Study} \label{sec: example}

The ASSISTments platform is an educational technology used for teaching math in schools. The online platform provides teachers with tools to deliver formative assessments, allowing students to receive immediate feedback as they work through problems. Teachers use the platform to monitor student performance and growth, allowing them to adjust their classroom instruction to match the knowledge base of the class. The platform has been evaluated in two large efficacy studies in schools, one in Maine \parencite{rochelle_how_2017} and the other in North Carolina \parencite{feng_implementing_2023}. Each of these studies was designed to estimate and test hypotheses regarding the average treatment effect. For this example, we focus on the North Carolina study, using it to illustrate how results from this paper could be used when designing a study.

The evaluation took place in North Carolina because it was identified as a state that is "more geographically representative of the U.S." than the previous study in Maine. The paper provides a comparison of the population of schools in North Carolina to those in the U.S. on five covariates: Percent of White students, percent of Hispanic students, percent of Black students, Percent of students receiving free or reduced priced lunch (a proxy for socioeconomic status), and the Percent of schools that were rural. In the pre-registration plan (https://sreereg.icpsr.umich.edu/framework/pdf/index.php?id=2064), the study design focuses on $N = 80$ schools, with equal proportions randomized to the treatment or business as usual. The pre-registration also assumes a pre-test measure would be used and would explain 80 percent of the variation in outcomes. 

In addition to identifying a target population ($P_B$) --- schools serving 7th grade students in the U.S. --- the evaluation also provides us with some of the design parameters needed to explore predictive models: the total sample size planned ($N = 80$); the  number of covariates potentially of interest ($p = 5$); a school level pretest covariate highly related to the outcome ($R_0^2 = .8$). This leaves three parameters without values: the standardized treatment effect variation ($\tau_*^2$), the correlation between Y(0) and treatment effects ($\tau_{0\eta}$), and the proportion of variation in treatment effects explainable by $p$ covariates ($R_{\tau p}^2$). The latter two of these parameters are very rarely reported in studies, making it difficult to anticipate their values in practice. For these we will use a sensitivity approach.

Fortunately, there is some information regarding prior degrees of treatment effect variation. \textcite{weiss_how_2017} examine data from 16 multisite randomized trials. This includes five trials in early childhood and elementary schools, seven in middle and high schools, two in post-secondary education, and two in labor or workforce development; each study provided multiple outcomes. Their analysis includes exploring both average treatment effects found in these studies, as well as the degree of variation in site-average treatment effects. These parameter estimates are provided in their Table 4. Overall, the estimates of $\tau_*$ range from $0$ to $0.35$, with 90-percent upper confidence values of between $0$ to $0.49$. This indicates values of the standardized variance between $0$ and $.25$ SDs. For this analysis, we will consider three values of $\tau_*^2$: low ($.10^2$), medium ($.25^2$), and large ($.5^2$).

Unfortunately, this range of values is quite large. Further analyses by Weiss et al indicated that treatment effect variation was larger when the intervention had low specificity, low intensity, and when the comparison group was served in a different building or site. Since the ASSISTments program is a supplemental program and since schools are randomized, this suggests there may be a high degree of variation. Nonetheless, we will explore several values. For these explorations, we focus on the randomization of schools, assuming large enough samples of students within each school that we can ignore this nesting. Throughout we also assume equal treatment and comparison group sizes. 

\subsubsection{ANOVA}

From Equation \ref{eq:mdes}, we can see that the MDES with $N = 80$ schools and $R_0^2 = 0.80$ is $0.22$ SDs. In order to investigate prediction, we begin by assuming a model in which the treatment effect for every school in the target population ($P_B$) is predicted to be the same ($\hat{\delta_i} = \hat{\Delta_A}$), yet the true school treatment effects vary. 

Table \ref{tab:anovapred} explores the MSPE and 90-percent prediction interval width for three treatment group sizes ($n = 40, 100, 500$) and the three standardized treatment effect variation values defined above. This table shows that even with a small degree of variation, the width of a 90-percent prediction interval is quite large. For example, with an average effect size of $.22$, the 90-percent prediction intervals include both negative and positive values for nearly all cases except when the variation in treatment effects is small ($\tau_*^2 = .1^2$). Thus, even though the sample size is adequate for testing if on average the effect of ASSISTments is zero, if treatment effect variation is moderate or large, the estimate of this effect will not be adequately sensitive for predicting if ASSISTments would work in any particular school. 

\begin{table}[h]
\centering
\begin{tabular}{@{}lllllll@{}}
\toprule
n & p & $\tau_*^2$ & $MSPE_p$ & 90PI W & 90PI LB & 9PI UB \\ \midrule
40 & 1 & 0.01 & 0.018 & 0.438 & 0.001 & 0.439 \\
40 & 1 & 0.0625 & 0.071 & 0.878 & -0.219 & 0.659 \\
40 & 1 & 0.25 & 0.262 & 1.685 & -0.622 & 1.062 \\ \midrule
40 & 3 & 0.01 & 0.023 & 0.497 & -0.028 & 0.468 \\
40 & 3 & 0.0625 & 0.077 & 0.913 & -0.236 & 0.676 \\
40 & 3 & 0.25 & 0.270 & 1.711 & -0.635 & 1.075 \\ \midrule
100 & 1 & 0.01 & 0.013 & 0.376 & 0.032 & 0.408 \\
100 & 1 & 0.0625 & 0.066 & 0.845 & -0.203 & 0.643 \\
100 & 1 & 0.25 & 0.255 & 1.661 & -0.610 & 1.050 \\ \midrule
100 & 3 & 0.01 & 0.015 & 0.405 & 0.018 & 0.422 \\
100 & 3 & 0.0625 & 0.068 & 0.860 & -0.210 & 0.650 \\
100 & 3 & 0.25 & 0.258 & 1.672 & -0.616 & 1.056 \\ \midrule
500 & 1 & 0.01 & 0.011 & 0.339 & 0.051 & 0.389 \\
500 & 1 & 0.0625 & 0.063 & 0.827 & -0.194 & 0.634 \\
500 & 1 & 0.25 & 0.251 & 1.648 & -0.604 & 1.044 \\ \midrule
500 & 3 & 0.01 & 0.011 & 0.345 & 0.047 & 0.393 \\
500 & 3 & 0.0625 & 0.064 & 0.830 & -0.195 & 0.635 \\
500 & 3 & 0.25 & 0.252 & 1.650 & -0.605 & 1.045 \\ \bottomrule
\end{tabular}
\caption{\emph{For all computations, $\rho_{0\eta} = 0$ and $R_0^2 = .80$ assumed. Total sample size for a study is $N=2n$.}}
\label{tab:anovapred}
\end{table}

\subsubsection{Moderator Model}

Given the width of these prediction intervals, a question a researcher might have would be if such intervals could be made smaller by using a moderator model. Applying Equation \ref{eq:minR2}, in Table \ref{tab:minR2} we provide the minimum $R_{\tau p}^2$ value required in order for predictive model using $p = 1$ or $3$ moderators to outperform an ANCOVA model (i.e., lower MSPE). The table indicates that if the degree of variation in the effect of ASSISTments across schools is small, in a sample with $2n = 80$ schools, a school-specific treatment effect prediction only outperforms the average treatment effect if the moderator explains \emph{all} of the variation in treatment effects ($R_\tau^2 = 1$). In comparison, with a sample of size $2n = 200$, the moderator model is preferable if the moderator explains only 50 percent of the variation in treatment effects. 

If the variation in the effects of ASSISTments was expected to be larger --- as we have assumed here --- then the ANCOVA model is not always best. For example, with a sample of $2n = 80$ and moderate treatment effect variation, a single moderator would need to be able to explain at least $22$ percent of the variation in treatment effects; if large, then only $8$ percent of the variation. Importantly, note that if a very large study were possible (as in trials with individual random assignment), these requirements drop substantially. For example, with $2n = 1,000$, a moderator would only need to be able to explain $<2$-percent of the treatment effect variation to improve the MSPE.

\begin{table}[h]
\centering
\begin{tabular}{@{}llll@{}}
\toprule
n & p & $\tau_*^2$ & $min(R_\tau^2)$ \\ \midrule
40 & 1 & 0.01 & 100\% \\
40 & 1 & 0.0625 & 22\% \\
40 & 1 & 0.25 & 8\% \\ \midrule
40 & 3 & 0.01 & 100\% \\
40 & 3 & 0.0625 & 46\% \\
40 & 3 & 0.25 & 16\% \\ \midrule
100 & 1 & 0.01 & 50\% \\
100 & 1 & 0.0625 & 9\% \\
100 & 1 & 0.25 & 3\% \\ \midrule
100 & 3 & 0.01 & 100\% \\
100 & 3 & 0.0625 & 20\% \\
100 & 3 & 0.25 & 7\% \\ \midrule
500 & 1 & 0.01 & 10\% \\
500 & 1 & 0.0625 & 2\% \\
500 & 1 & 0.25 & 1\% \\ \midrule
500 & 3 & 0.01 & 22\% \\
500 & 3 & 0.0625 & 4\% \\
500 & 3 & 0.25 & 1\% \\ \bottomrule
\end{tabular}
\caption{\emph{For all computations, $\rho_{0\eta} = 0$ and $R_0^2 = .80$ assumed. Total sample size for a study is $N=2n$. $min(R_\tau^2)$ is the minimum value of $R_\tau^2$ required for $MSPE_{2p} < MSPE_p$.}}
\label{tab:minR2}
\end{table}

If such a moderator (or three) were available, how much might their inclusion reduce the predictive error? In Table \ref{tab:abserror} we investigate this further. For this table, we include values of $R_{\tau p}^2$ in increments of $0.20$, and include only rows in which the predictive model with a single moderator outperforms the average treatment effect (constant) model (see Appendix B for values with $p=3$ moderators). For each row, we include the width of a 90-percent prediction interval for each model; the final column indicates how much smaller this interval is for different values of $R_\tau^2$.  Reading from this table, for example, we see that if the variation in the effect of ASSISTments is moderate, a covariate explaining $40$ percent of the treatment effects would reduce the prediction interval by about $9$ percent, while one explaining $80$ percent could reduce the interval by about $32$ percent. In absolute terms, however, with a sample size of $N=80$ schools, even with a strong predictor, these prediction intervals are still very wide. Overall, this means that while moderators can improve the accuracy of predictions, to do so they need to be either very strongly predictive of treatment effects or sample sizes need to be substantially larger than typical in cluster randomized studies. 

\begin{table}[]
\centering
\begin{tabular}{@{}lllllllll@{}}
\toprule
\cellcolor[HTML]{FFFFFF}p & n & $R_\tau^2$ & $\tau_*^2$ & $MSPE_p$ & $MSPE_{2p}$ & 90PIW(p) & 90PIW(2p) & PctRedW \\ \midrule
1 & 40 & 0.4 & 0.0625 & 0.071 & 0.059 & 0.88 & 0.80 & 9\% \\
1 & 40 & 0.6 & 0.0625 & 0.071 & 0.046 & 0.88 & 0.71 & 19\% \\
1 & 40 & 0.8 & 0.0625 & 0.071 & 0.033 & 0.88 & 0.60 & 32\% \\
1 & 40 & 1 & 0.0625 & 0.071 & 0.020 & 0.88 & \cellcolor[HTML]{EFEFEF}0.47 & 47\% \\ \midrule
1 & 40 & 0.2 & 0.25 & 0.262 & 0.230 & 1.68 & 1.58 & 6\% \\
1 & 40 & 0.4 & 0.25 & 0.262 & 0.178 & 1.68 & 1.39 & 18\% \\
1 & 40 & 0.6 & 0.25 & 0.262 & 0.125 & 1.68 & 1.16 & 31\% \\
1 & 40 & 0.8 & 0.25 & 0.262 & 0.073 & 1.68 & 0.89 & 47\% \\
1 & 40 & 1 & 0.25 & 0.262 & 0.020 & 1.68 & \cellcolor[HTML]{EFEFEF}0.47 & 72\% \\ \midrule
1 & 100 & 0.6 & 0.01 & 0.013 & 0.012 & 0.38 & \cellcolor[HTML]{EFEFEF}0.36 & 4\% \\
1 & 100 & 0.8 & 0.01 & 0.013 & 0.010 & 0.38 & \cellcolor[HTML]{EFEFEF}0.33 & 12\% \\
1 & 100 & 1 & 0.01 & 0.013 & 0.008 & 0.38 & \cellcolor[HTML]{EFEFEF}0.29 & 22\% \\ \midrule
1 & 100 & 0.2 & 0.0625 & 0.066 & 0.059 & 0.85 & 0.80 & 5\% \\
1 & 100 & 0.4 & 0.0625 & 0.066 & 0.046 & 0.85 & 0.71 & 16\% \\
1 & 100 & 0.6 & 0.0625 & 0.066 & 0.034 & 0.85 & 0.60 & 29\% \\
1 & 100 & 0.8 & 0.0625 & 0.066 & 0.021 & 0.85 & \cellcolor[HTML]{EFEFEF}0.47 & 44\% \\
1 & 100 & 1 & 0.0625 & 0.066 & 0.008 & 0.85 & \cellcolor[HTML]{EFEFEF}0.29 & 65\% \\ \midrule
1 & 100 & 0.2 & 0.25 & 0.255 & 0.212 & 1.66 & 1.51 & 9\% \\
1 & 100 & 0.4 & 0.25 & 0.255 & 0.161 & 1.66 & 1.32 & 21\% \\
1 & 100 & 0.6 & 0.25 & 0.255 & 0.110 & 1.66 & 1.09 & 34\% \\
1 & 100 & 0.8 & 0.25 & 0.255 & 0.059 & 1.66 & 0.80 & 52\% \\
1 & 100 & 1 & 0.25 & 0.255 & 0.008 & 1.66 & \cellcolor[HTML]{EFEFEF}0.29 & 82\% \\ \midrule
1 & 500 & 0.2 & 0.01 & 0.011 & 0.010 & 0.34 & \cellcolor[HTML]{EFEFEF}0.32 & 5\% \\
1 & 500 & 0.4 & 0.01 & 0.011 & 0.008 & 0.34 & \cellcolor[HTML]{EFEFEF}0.29 & 15\% \\
1 & 500 & 0.6 & 0.01 & 0.011 & 0.006 & 0.34 & \cellcolor[HTML]{EFEFEF}0.25 & 27\% \\
1 & 500 & 0.8 & 0.01 & 0.011 & 0.004 & 0.34 & \cellcolor[HTML]{EFEFEF}0.20 & 42\% \\
1 & 500 & 1 & 0.01 & 0.011 & 0.002 & 0.34 & \cellcolor[HTML]{EFEFEF}0.13 & 61\% \\ \midrule
1 & 500 & 0.2 & 0.0625 & 0.063 & 0.052 & 0.83 & 0.75 & 9\% \\
1 & 500 & 0.4 & 0.0625 & 0.063 & 0.039 & 0.83 & 0.65 & 21\% \\
1 & 500 & 0.6 & 0.0625 & 0.063 & 0.027 & 0.83 & 0.54 & 35\% \\
1 & 500 & 0.8 & 0.0625 & 0.063 & 0.014 & 0.83 & \cellcolor[HTML]{EFEFEF}0.39 & 53\% \\
1 & 500 & 1 & 0.0625 & 0.063 & 0.002 & 0.83 & \cellcolor[HTML]{EFEFEF}0.13 & 84\% \\ \midrule
1 & 500 & 0.2 & 0.25 & 0.251 & 0.202 & 1.65 & 1.48 & 10\% \\
1 & 500 & 0.4 & 0.25 & 0.251 & 0.152 & 1.65 & 1.28 & 22\% \\
1 & 500 & 0.6 & 0.25 & 0.251 & 0.102 & 1.65 & 1.05 & 36\% \\
1 & 500 & 0.8 & 0.25 & 0.251 & 0.052 & 1.65 & 0.75 & 55\% \\
1 & 500 & 1 & 0.25 & 0.251 & 0.002 & 1.65 & \cellcolor[HTML]{EFEFEF}0.13 & 92\% \\ \bottomrule
\end{tabular}
\caption{\emph{For all computations, $\rho_{0\eta} = 0$ and $R_0^2 = .80$ assumed. Total sample size for a study is $N=2n$.Gray highlights are 90PI widths smaller than 0.50.}}
\label{tab:abserror}
\end{table}

\subsubsection{Population choices}
In the ASSISTments study, the goal was to predict treatment effects for all public schools serving 7th graders in the U.S. However, the study itself only recruited schools in North Carolina. A question then is what penalty could be exerted by conducting the study in one population ($P_A$) but predicting treatment effects in another ($P_B$).

To investigate this, we returned to the Common Core of Data and defined the target population as non-charter, non-virtual, public U.S. schools serving at least $30$ 7th graders. This resulted in a population of 16,775 schools. We then limited the population to the subset of these schools that had covariate data on all five identified variables --- percent White; percent Hispanic; percent Black; percent free-or-reduced price lunch (an indicator of low socioeconomic status); and Rural. The percent of students receiving FRL was not reported for 484 schools; this included all 400 schools in Massachusetts. The final target population $P_B$ thus included 16,290 schools in the United States. Of these, 536 schools are in North Carolina, the population where the study took place ($P_A$).

We began by comparing the two populations using metrics common in the generalization literature (where the focus is on estimation of the average treatment effect). One metric is the absolute SMD, while the other is the variance ratio; in both cases, it is typical to standardize with respect to the target population ($P_B$). Here the absolute SMDs between the two groups range from 0.14SD to 0.42SD, with three of these larger than the commonly used 0.25SD threshold. The variance ratios ($P_A$ to in $P_B$) range between 0.24 and 1.19, with only one outside the threshold of 0.5 to 2. The generalizability index for these two groups is $0.95$, indicating that the population of schools in NC ($P_A$) is nearly as similar to those in the US ($P_B$) as a random sample. Altogether this suggests that the estimate of the ATE could be generalized to the target population ATE easily.

In this paper, we have shown that when considering prediction, the standardization that matters for MSPE is with respect to the sampled population $P_A$ not the target population $P_B$. On these five covariates, the Mahalanbois distance between the two populations is $M = 0.55$, while the Burg distance is $9.71$. When $P_A \equiv P_B$, we would expect $M+B = p = 5$; here instead $M+B = 9.71$, about twice as large. This suggests that while the two populations are similar \emph{on average}, there is generally less variation across these covariates in $P_A$ than in $P_B$. If left unadjusted, this would result in a MSPE that is about \emph{twice} as large as if $P_A \equiv P_B$.

To adjust for these differences, inverse odds weights could be used. Here we predicted the outcome $Z$, where $Z = 1$ if a school was in North Carolina and $Z = 0$ if it was in the target population. We used a logistic regression model and included the five covariates identified before. Based on this, we calculated inverse odds weights using Equation \ref{eq: inverseodds}. We then compared the distribution of these weights in North Carolina versus in the US, with a focus on identifying schools in the common support of these two distributions. We found that $98.3$ percent of schools in the US were within the range of weights identified in North Carolina. Thus, in order to meet the positivity assumption (A2), we restrict the target population to this slightly smaller subset. In practice, this would mean that the resulting model could be used to predict treatment effects for all but $1.7$ percent of US public schools serving 7th graders.

Finally, we calculated the variance inflation penalty that would be incurred as a result of these weights (Equation \ref{eq:vif}). This penalty was found to be $VIF = 1.42$, indicating that the actual MSPE would be about $42$ percent larger as a result of this reweighting. Importantly, while this is smaller than the doubling expected without adjustment (from the $M+B$ versus $p$ in the analysis above), it still exerts a large penalty. This would mean that by limiting the sample of schools to those in North Carolina ($P_A$), a sample of $N*M_{VIF} = 113$ schools would be required to have the same accuracy as a sample of $N = 80$ schools in the US ($P_B$).

\section{Conclusion} \label{sec: conclusion}

In this paper, we have examined the conditions under which prediction of unit-specific treatment effects is possible on the basis of results from a randomized trial. We focus in particular on the development of intuition and functions that can be useful for planning studies. To develop these intuitions, we have focused on linear parametric models, as they provide closed form results. 

\textbf{For those designing and conducting experiments}, it is easy to focus on the development and the use of a predictive model without thinking carefully about its performance. Those working with RCTs are often aware of a variety of rules of thumb related to statistical power and sensitivity, all of which have to do with the ATE. We have shown, however, that these  rules of thumb do not directly transfer to the predictive case. For example, predictive error involves new parameters, the trickiest of which is the degree of idiosyncratic variation. As we have shown, this is a function of a completely unknowable parameter -- the correlation between potential outcomes. The only information truly available in data here is the degree to which the residual variances in the two groups (T = 0, 1) is the same. If they differ, then this idiosyncratic variation is clearly non-zero. But if they are the same, this does not prove that there is no idiosyncratic variation. To some extent, our choice for the consideration of the value of this correlation $\rho_{01|\bm{x}}$ must depend upon assumptions regarding the very nature of treatment effects: do we think that even under ideal circumstances they would follow a pattern that could be predicted? Or is there some part of them that is truly idiosyncratic --- times when treatments happen to work for some for reasons that are truly random?

Regardless, we have provided formulas that can be used to determine the types of moderators and sample sizes that are needed to provide accurate predictions of unit-specific treatment effects. We have shown that in the small sample sizes found in cluster-randomized trials --- where predictions of site specific treatment effects or other aggregates are desired --- the ATE is often the most accurate prediction of unit specific treatment effects. However, when the ATE is small, unless the treatment effect variation is also small, the resulting prediction may not be adequate for distinguishing between units with positive or negative treatment effects (i.e., prediction intervals include zero).  We have also shown that in order to outperform the ATE -- thus providing different predicted treatment effects for different units -- the moderators included need to be highly predictive of treatment effects. If this is not the case, then larger sample sizes are required. 

\textbf{Methods for quantifying predictive error} have long existed in OLS regression; for example, we know that these predictive errors are larger, since they involve the residual from a new observation. As we have shown here, however, these formulas are too simple once we move out of the $P_A$ case. That is, the formulas and estimators are only valid if our sample is a random sample from the population. When the sample might be highly selected (e.g., a combination of convenience and eligibility criteria) -- as is typically the case in RCTs -- the errors involve additional components. One of these components involves the introduction of bias that may arise from differences in the support for the covariates in different populations, called ``distribution shift'' in the language of ML. Other differences have to do with the degree of similarity between the means, variances, and covariances of these covariates in the two populations. Perhaps what is hardest here is that for a given sample, again it may be difficult to accurately quantify these terms. Thus, by all metrics available and calculable with the data, the model may appear to be performing well --- and yet not perform well at all in the target population. 

\textbf{The findings here are very much related to those in the generalizability literature}, though they differ in important ways as well. The literature on generalizability has focused strongly on estimation of the average treatment effect in one or more target populations. These findings suggest, for example, that if one wants to design a study that minimizes this bias, they should match the first moments of the sample to the target population. Here we find that if the goal is prediction, this extends further --- matching the means alone is simply not enough. Instead, we need to match the variances of the moderators as well. Importantly, however, we also show that an approach that further reduces the error (MSPE) is to purposely select the sample so that it maximizes heterogeneity in the covariates --- that doing so can reduce the MSPE even beyond that of a random sample. 

Importantly, there is another difference. In the literature on generalizability, the focus has been on bias in the estimate of the ATE. While it can be counterintuitive, even when the ATE estimate is biased, there is no bias in its associated standard error. This is because the standard error has to do with the sampling variation in the data generating process, which focuses on the past. But in prediction, the focus is on the future. This means that the standard estimators of the predictive error --- e.g., based upon the observed variation in residuals in the sample --- can also be biased. Again, this requires adjusting not only the predictions themselves, but also their measures of error.

Finally, our analysis suggests that if one is planning a study with prediction in mind, the broadest possible target population should be anticipated. As we have shown, it is simply not possible to build a strong predictive model of treatment effect heterogeneity without heterogeneity in the covariates and outcome. Put another way, we need heterogeneity ``in'' in order to get heterogeneity ``out''. Ultimately, this means that while heterogeneity is often seen as our enemy when estimating the ATE, in prediction heterogeneity is our friend.

\printbibliography

\appendix

\section{Appendix: Proofs}

%MSPE Equation 3.20, for planning any model w/ mod
\subsection{Proof of Equation \ref{eq:mspe_full}}
First, write
\begin{align}
    Y_i(0) &= \mu_{0} + x_i'\bm{\beta}_0 + \epsilon_{i0} \\
    Y_i(1) &= \mu_{1} + x_i'\bm{\beta}_1 + \epsilon_{i1}  \\
    &= (\mu_{0} + \Delta_A) + x_i'\bm{\delta} + \eta_i + \epsilon_{i0}
\end{align}

Then $V[Y_i(0)] = \sigma_{0}^2$ and $V[Y_i(1)] = \sigma_{0}^2 + \tau^2 + 2\rho_{0\eta}\sigma_0\tau $. Let $ R_{-0}^2 = 1 - R_{0}^2 = \sigma_{0|x}^2/\sigma_0^2 $ and $ R_{-\tau}^2 = 1 - R_\tau^2 = \tau_{A|x}^2 / \tau_{A}^2 $. Let $ \tau_{*}^2 = \tau_A^2/ \sigma_0^2 $. Now, via substitution we have:

\begin{align}
MSPE(\hat{\delta_i}) & = \left(\frac{\sigma_{0|x}^2}{n} + \frac{\sigma_{1|x}^2}{n} \right)(1+p) + \tau_{A|x}^2 \\
    & = \left(\frac{1+p}{n} \right) \left( \sigma_{0|x}^2 + (\sigma_{0|x}^2 + \tau_{A|x}^2 + 2\rho_{0\eta|x}\sigma_{0|x}\tau_{A|x})  \right) + \tau_{A|x}^2 \\
    & = \sigma_0^2 \left(\frac{1+p}{n} \right) \left( 2 R_{-0}^2 + R_{-\tau}^2\tau_{*}^2 + 2\rho_{0\eta|x}\tau_{*|x}R_{-0}R_{-\tau}  \right) + R_{-\tau}^2\tau_{A}^2 \\
    & = 2\sigma_0^2 \left(\frac{1+p}{n} \right) \left[ R_{-0}^2 + \rho_{0\eta|x}\tau_{*|x}R_{-0}R_{-\tau} + \tau_{*}^2R_{-\tau}^2 \left(\frac{1}{2} + \frac{n}{2(1+p)} \right) \right]
\end{align}

%MSPE Equation 3.23, ancova planning
\subsection{Proof of Equation \ref{eq:mspe_ancova}}

First, recall that we will estimate the model,
\begin{equation}
    Y_i = \mu_0 + \Delta T_i + \bm{x}_i'\bm{\beta} + \epsilon_i
\end{equation}
If we now split the data into the two groups, we have:
\begin{align}
    Y_i(0) & = \mu_0 + \bm{x}_i'\bm{\beta} + \epsilon_{i0} \\
    Y_i(1) & = (\mu_0 + \Delta) + \bm{x}_i'\bm{\beta} + \epsilon_{i1}
\end{align}
Now, assume we have two groups, each with sample size $n$. We estimate $V(\epsilon_i) = \sigma^2$ using a pooled estimator, with $s^2 = (s_0^2 + s_1^2)/2$. Thus, we have $\sigma^2 = E(s^2) = (\sigma_0^2 + \sigma_1^2)/2$.  Applying results from the proof of Equation \ref{eq:mspe_full}, we have
\begin{align}
    \sigma_{|x}^2 & = (\sigma_{0|x}^2 + \sigma_{1|x}^2)/2  \\
    & = \left( 2\sigma_{0|x}^2 + \tau_{A|x}^2 + 2\rho_{0\eta|x}\sigma_{0|x}\tau_{A|x} \right)/2
\end{align}

Now, by substitution and rearrangement we have:
\begin{align} 
MSPE(\hat{\delta_i}|ANCOVA) & = \frac{\sigma^2(2+p)R_{-}^2}{2n} + \tau_{A}^2 \\
& = \left(\frac{2+p}{4n} \right) (2\sigma_{0|x}^2 + \tau_A^2 + 2\rho_{0\eta|x}\tau_A\sigma_{0|x} ) + \tau^2 \\
& = \sigma_0^2 \left[ ( 2R_{-0}^2 + 2\rho_{0\eta|x}\tau_{*}R_{-0}) \left(\frac{2+p}{2n} \right) + \tau_*^2 \left( \frac{2+p}{4n} + 1 \right) \right] \\
& = \sigma_0^2 \left(\frac{2+p}{2n} \right) \left[ R_{-0}^2 + \rho_{0\eta}\tau_*R_{-0} + \tau_*^2 \left( \frac{1}{2} + \frac{2n}{2+p} \right)\right]
\end{align}

Finally, note that the raw means is a special case of this ANCOVA result, substituting $R_{-0}^2 = 1$ and $p = 0$.

%R2 minimum for preferring ancova, eq 3.26
\subsection{Proof of Equation \ref{eq:minR2}}

We wish to find the value of $R_{\tau}^2 = r_{\tau}^2$ such that when $R_{\tau}^2 < r_{\tau}^2$ we have $MSPE(ANCOVA) < MSPE(Mod)$ where the ANCOVA estimator includes $2+p$ parameters and the Moderator estimator includes $2(1+p)$ parameters. 

To do so, first we write and rearrange:
\begin{align}
MSPE[2+p] & < MSPE[2(1+p)] \\
n MSPE[2+p] &< 2(1+p)\left[R_{-0}^2 + \rho_{0\eta|x}\tau_*R_{-0}R_{-\tau} + \tau_*^2R_{-\tau}\left(\frac{1}{2} + \frac{n}{2(1+p)} \right) \right] \\
0 &< R_{-\tau}^2A + R_{-\tau}B + C 
\end{align}
where now
\begin{align}
A &= \tau_*^2\left(\frac{1}{2} + \frac{n}{2(1+p)} \right) \\
B &= 2(1+p)\rho_{0\eta|x}\tau_*R_{-0} \\
C &= 2(1+p)R_{-0}^2 - nMSPE[2+p]
\end{align}

Thus we need to solve the quadratic equation, with 
\begin{align}
R_{-\tau} = \frac{-B \pm \sqrt{B^2 - 4AC}}{2A}
\end{align}
The final result falls from substitution and rearrangement, solving for $R_\tau^2 = 1 - R_{-\tau}^2$. Note that when there is no real root (because $B^2 - 4AC < 0$), the ANOVA model is always preferred.

\section{Appendix: p = 3 table}
This table provides a supplement to Table \ref{tab:abserror}, now focused on $p = 3$ covariates.
% \usepackage{booktabs}
\begin{table}[]
\centering
\begin{tabular}{@{}lllllllll@{}}
\toprule
$p$ & $n$ & $R_{\tau}^2$ & $\tau_{s}^2$ & $MSPE_{p}$ & $MSPE_{2p}$ & $90PIW(p)$ & $90PIW(2p)$ & PctRedW \\ \midrule
3 & 40 & 0.6 & 0.0625 & 0.077 & 0.068 & 0.91 & 0.85 & 6\% \\
3 & 40 & 0.8 & 0.0625 & 0.077 & 0.054 & 0.91 & 0.76 & 16\% \\
3 & 40 & 1 & 0.0625 & 0.077 & 0.040 & 0.91 & 0.66 & 28\% \\ \midrule
3 & 40 & 0.2 & 0.25 & 0.270 & 0.260 & 1.71 & 1.68 & 2\% \\
3 & 40 & 0.4 & 0.25 & 0.270 & 0.205 & 1.71 & 1.49 & 13\% \\
3 & 40 & 0.6 & 0.25 & 0.270 & 0.150 & 1.71 & 1.27 & 26\% \\
3 & 40 & 0.8 & 0.25 & 0.270 & 0.095 & 1.71 & 1.01 & 41\% \\
3 & 40 & 1 & 0.25 & 0.270 & 0.040 & 1.71 & 0.66 & 62\% \\ \midrule
3 & 100 & 0.2 & 0.0625 & 0.068 & 0.068 & 0.86 & 0.86 & 0\% \\
3 & 100 & 0.4 & 0.0625 & 0.068 & 0.055 & 0.86 & 0.77 & 10\% \\
3 & 100 & 0.6 & 0.0625 & 0.068 & 0.042 & 0.86 & 0.67 & 22\% \\
3 & 100 & 0.8 & 0.0625 & 0.068 & 0.029 & 0.86 & 0.56 & 35\% \\
3 & 100 & 1 & 0.0625 & 0.068 & 0.016 & 0.86 & 0.42 & 52\% \\ \midrule
3 & 100 & 0.2 & 0.25 & 0.258 & 0.224 & 1.67 & 1.56 & 7\% \\
3 & 100 & 0.4 & 0.25 & 0.258 & 0.172 & 1.67 & 1.36 & 18\% \\
3 & 100 & 0.6 & 0.25 & 0.258 & 0.120 & 1.67 & 1.14 & 32\% \\
3 & 100 & 0.8 & 0.25 & 0.258 & 0.068 & 1.67 & 0.86 & 49\% \\
3 & 100 & 1 & 0.25 & 0.258 & 0.016 & 1.67 & 0.42 & 75\% \\ \midrule
3 & 500 & 0.4 & 0.01 & 0.011 & 0.009 & 0.35 & 0.32 & 8\% \\
3 & 500 & 0.6 & 0.01 & 0.011 & 0.007 & 0.35 & 0.28 & 19\% \\
3 & 500 & 0.8 & 0.01 & 0.011 & 0.005 & 0.35 & 0.24 & 31\% \\
3 & 500 & 1 & 0.01 & 0.011 & 0.003 & 0.35 & 0.19 & 46\% \\ \midrule
3 & 500 & 0.2 & 0.0625 & 0.064 & 0.054 & 0.83 & 0.76 & 8\% \\
3 & 500 & 0.4 & 0.0625 & 0.064 & 0.041 & 0.83 & 0.67 & 20\% \\
3 & 500 & 0.6 & 0.0625 & 0.064 & 0.028 & 0.83 & 0.55 & 33\% \\
3 & 500 & 0.8 & 0.0625 & 0.064 & 0.016 & 0.83 & 0.41 & 50\% \\
3 & 500 & 1 & 0.0625 & 0.064 & 0.003 & 0.83 & 0.19 & 78\% \\ \midrule
3 & 500 & 0.2 & 0.25 & 0.252 & 0.205 & 1.65 & 1.49 & 10\% \\
3 & 500 & 0.4 & 0.25 & 0.252 & 0.154 & 1.65 & 1.29 & 22\% \\
3 & 500 & 0.6 & 0.25 & 0.252 & 0.104 & 1.65 & 1.06 & 36\% \\
3 & 500 & 0.8 & 0.25 & 0.252 & 0.054 & 1.65 & 0.76 & 54\% \\
3 & 500 & 1 & 0.25 & 0.252 & 0.003 & 1.65 & 0.19 & 89\% \\ \bottomrule
\end{tabular}
\caption{For all computations, $\rho_{0\eta} = 0$ and $R_0^2 = .80$ assumed. Total sample size for a study is $N=2n$.}
\label{tab:my-table}
\end{table}

\end{document}